\documentclass[aps,twocolumn,showpacs,floatfix,preprintnumbers]{revtex4}
\begin{document}
\preprint{CAPCyR-FCFM-BUAP-01-05}
\preprint{LBNL-57358}
\title{Solving the SUSY CP problem with flavor breaking F-terms}
\author{J. Lorenzo Diaz-Cruz}
\email{lorenzo.diaz@fcfm.buap.mx}
\affiliation{Cuerpo Acad\'emico de Part\'\i culas, Campos y Relatividad FCFM \\
BUAP,  Ap.  Postal 1364 Puebla Pue., 72000, M\'exico}
\author{Javier Ferrandis}
\email{ferrandis@mac.com}
\homepage{http://homepage.mac.com/ferrandis}
\affiliation{MEC postdoctoral fellow at the Theoretical Physics Group\\ 
Lawrence Berkeley National Laboratory  \\
Cyclotron Road 1, Berkeley CA 94720}
\begin{abstract}
Supersymmetric flavor models 
for the radiative generation of fermion masses offer an
alternative way to solve the SUSY-CP problem. 
We assume that the supersymmetric theory
is flavor and CP conserving.
CP violating phases are
associated to the vacuum expectation values of flavor violating susy-breaking fields. 
As a consequence, phases appear at tree level
only in the soft supersymmetry breaking matrices.
Using a $U(2)$ flavor model as an example
we show that it is possible to generate radiatively the first and second generation of
quark masses and mixings as well as the CKM CP phase. The one-loop
supersymmetric contributions to EDMs are automatically zero
since all the relevant parameters in the lagrangian are 
flavor conserving and as a consequence real.
The size of the flavor and CP mixing in the susy breaking sector
is mostly determined by the fermion mass ratios and CKM elements. 
We calculate the contributions to 
$\epsilon$, $\epsilon^{\prime}$ and to the CP asymmetries in the B
decays to $\psi K_{s}, \phi K_{s}, \eta^{\prime} K_{s}$ and $X_{s}\gamma$.
We analyze a case study with maximal predictivity in the
fermion sector. For this worst case scenario the measurements of 
$\Delta m_{K}$,$\Delta m_{B}$ and $\epsilon$ constrain the model 
requiring extremely heavy squark spectra.
 \end{abstract}
\maketitle
\newpage
%
\section{Introduction}
It was suggested a few years ago that the huge number of 
possible string theory vacua in combination with 
eternal inflation may allow us to understand the smallness of the cosmological constant 
from an anthropic point of view \cite{Bousso:2000xa,Susskind:2003kw}.
Although there is no current general framework for examining these metastable 
vacua in string theory \cite{Banks:2003es} some particular 
methods have been proposed \cite{landmethods}.
It is still an open debate whether the landscape 
does or does not predict high scale supersymmetry (SUSY) \cite{Dine:2004is,Susskind:2004uv}. 
Statistical analysis of the vacuum of certain string theories 
have derived formulae for the distribution of vacua, 
which support the idea
of a very high scale of supersymmetry breaking \cite{landstatistics}.  
It has been pointed out that this considerations may also be relevant
to understand the gauge hierarchy problem \cite{split}. 
This has motivated the recent interest in
field-theoretic realizations of models with large numbers of vacua \cite{fieldlandscape}
as well as in the analysis of the collider,
cosmological and other phenomenological implications
of supersymmetric models with very heavy supersymmetric spectra \cite{splitmore}.

In this paper we would like to revisit,
under the light of these new considerations, certain supersymmetric flavor models that 
are not usually considered in the literature because it is naively expected that 
they require a very heavy supersymmetric spectra to be compatible with experimental
constraints on flavor changing processes (FCNC). In particular we will show that 
supersymmetric flavor models for the radiative generation of fermion masses
generate very predictive Yukawa textures and at the same time offer a
new insight in the SUSY CP and flavor problems.

After more than 30 years of its observation, the violation of CP
symmetry is poorly known in its origin in the present particle physics paradigm,
despite its relevance in nature. 
The presence of CP violating phases in 
particle physics models can be tested, for instance,
through precision measurements of the electric dipole moments 
(EDMs) in the leptonic sector (electron and muon) and in the
quark sector (neutron and deuteron). At present there
are very stringent upper limits \cite{Eidelman:2004wy}, 
which are expected to improve by several orders of magnitude in the
near future experiments.
It is known that the Standard Model contribution
to the neutron EDM, in the absence of a theta term, is approximately $10^{-30}$~e~cm,
which is still more than four orders of magnitude below the reach 
of the current experiments. 

In the context of an unconstrained minimal supersymmetric
standard model (MSSM) \cite{CPreviews}
the generic contribution to the electric dipole moments
is several orders of magnitude larger than the SM contribution.
This serious volation of the current 
experimental constraints is known as the SUSY CP problem.  
Several possible explanations have been considered in the 
literature to account for the suppression of the supersymmetric contributions
to EDMs and other CP violating observables. Some of them are: 
1) CP suppression 2) cancellations, 3) alignment, 4) sfermion decoupling, 
and 5) flavor off-diagonal CP violation.  
\begin{itemize}
\item 
The CP suppressed scenario assumes that all the CP phases are suppressed \cite{Ellis:1982tk}
because CP is an approximate symmetry of the full theory.
\item 
The cancellation scenario is based on the existence of certain regions
of the SUSY parameter space where different contributions to EDMs
cancel  \cite{Ibrahim:1998je,nathibrahim}.
\end{itemize}
These two possibilities are nowadays ruled out.
The CP suppressed scenario would imply that all SM contributions to 
CP asymmetries are small, which we know today not to be the case
in the B system~\cite{CPsup}.
The cancellation scenario is also known to be ruled out
if constraints from electron, neutron and mercury atom EDMs are impossed
simultaneously \cite{Abel:2001vy}. Moreover there seems to be no 
symmetry that can guarantee such cancellations.
\begin{itemize}
\item 
In the CP alignment case
the phases associated to the relevant parameters 
are somehow related in such a way that the combinations 
contributing to EDMs cancel.
\item 
Decoupling entails that
the sfermion masses are heavy enough to strongly suppress 
the supersymmetric contributions even tough the CP phases can be 
arbitrarily large \cite{Nath:1991dn}.
\end{itemize}
The alignment scenario could arise naturally in the context of
models that generate exact soft universality as  
gauge mediated supersymmetry breaking models \cite{GMSB}
or in models with approximate horizontal abelian flavor symmetries \cite{alignment}.
The decoupling scenario is very plausible. It requires sfermion masses
of the order of several TeV, which implies the existence of 
fine tuning in the soft supersymmetry breaking sector.
\begin{itemize}
\item
The scenario with flavor off diagonal CP violation assumes that the 
origin of CP violation is closely related to the origin of flavor structures
in such a way that the flavor blind quantities as the $\mu$-term, soft bilinear terms
gaugino masses are real and only flavor off-diagonal CP phases are non-zero. 
\end{itemize}
We find that the
scenarios with flavor off-diagonal CP phases 
are especially interesting.
The models of this kind proposed in the literature to date \cite{Nir:1996am,Abel:2000hn,Ross:2004qn} 
assumed that flavor violating Yukawa matrices
and soft terms are both generated simultaneously at tree level 
at very high energies. Thus,
they require that Yukawa matrices and soft trilinear matrices are 
hermitian, which forces flavor-diagonal phases to vanish (up to
small RGE corrections). 
We would like to propose a 
new idea similar to the flavor off-diagonal scenario which
has not been considered before,  
\begin{itemize}
\item
We propose that the underlying supersymmetric theory is exactly CP conserving
while CP phases are only carried by flavor violating susy breaking fields.
\end{itemize}
At first sight one may be tempted to think that this scenario cannot
account for the observed CP violation in the SM,
especially the large mixing in the B-$\bar{\rm B}$ system,
and therefore conclude that CP violation must be present in 
the superpotential. We will show in this paper 
that this is not the case and certain models for the radiative
generation of first 
and second generation fermion masses and mixings recently proposed
\cite{Ferrandis:2004ri,Ferrandis:2004ng} allow us to generate
radiatively the CKM phase and 
offer an alternative solution for the SUSY CP problem.
  
In this model flavor and CP violation appear at tree level only in the soft supersymmetry
breaking parameters and are transmitted to the fermion sector at one loop through
low energy finite threshold corrections. CP violating phases could appear originally
in the vacuum expectation values of certain flavor violating susy breaking fields.
These vevs break spontaneously both flavor and the CP symmetry generating at tree level
flavor and CP violating soft mass matrices.
This class of models make use of the presence of soft supersymmetry
breaking terms for the radiative generation of quark and lepton masses through 
sfermion--gaugino loops, as originally suggested by 
W.~Buchmuller and D.~Wyler \cite{wyler1,wyler2} and later analyzed in more detail
in Refs.~\cite{Hall:1985dx,Banks:1987iu,Ma:1988fp,softRadSusy,Borzumati:1999sp,Arkani-Hamed:1995fq,su5relations}. The gaugino mass would provide the violation of fermionic chirality required by a fermion
mass while the soft breaking terms provide the violation of flavor and CP symmetries.

In this paper we have chosen to analyze a
case study that as we will show is the worst case scenario
from the point of view of FCNC constraints.
We analyze a model of this kind because it achieves
maximal predictivity in the quark sector. 
This case study, as we will see, tends to generate important contributions
to some flavor changing processes, especially $\Delta m_{K}$, $\Delta m_{d}$
and $\epsilon$, which can only be avoided if
both the squark and the gluino spectra are very heavy.

This paper is organized as follows. We begin in Sec.~\ref{model}
by describing the model we propose for the radiative generation
of first and second generation fermion masses and mixing angles.
In Sec.~\ref{RadYuk} we analyze the radiative generation of
Yukawa couplings in this model.
In Sec.~\ref{Radmasses} we study the predictions
and constraints for quark mass ratios. 
In Sec.~\ref{radCKM} we study the radiative generation
of the SM CKM phase and the predictions and constraints
arising from measured CKM elements and CP phases.
In Sec.\ref{EDMs} we argue that the
contributions to EDMs are exactly zero in this model. 
In Sec.~\ref{SCKM}
we analyze in detail the calculation of the soft matrices
in the SuperCKM basis. In Sec.~\ref{Ksystem}
we study the contributions to direct
and indirect CP violation in 
the K-$\bar{\rm K}$ system. In Sec.~\ref{Bsystem}
we study the contributions to CP asymmetries
in the B-$\bar{\rm B}$ system.
\section{The model \label{model}}
In this section we will consider a realistic three generation model
based in a horizontal ${\rm U(2)}_{H}$ symmetry \cite{bhrr}.
This is a generalization of the model 
proposed in Ref.~\cite{Ferrandis:2004ri}. 
We will assume the usual MSSM particle content where 
third generation matter superfields,
\begin{equation}
 {\cal Q}_{3}, {\cal D}_{3}, {\cal U}_{3}, {\cal L}_{3}, {\cal E}_{3},
\end{equation}
and up and down electroweak Higgs superfields,
${\cal H}_{u}$ and ${\cal H}_{d}$,
are singlets under ${\rm U(2)}_{H}$.
We will denote them abbreviately by $\bar{\phi}^{L}$ and $\phi^{R}$. Let us assume that
first and second generation left handed superfields,
\begin{equation}
\bar{\Psi}_{\cal Q} = \left(
 \begin{array}{c}
 {\cal Q}_{1}\\
{\cal Q}_{2}
\end{array}
\right),
\bar{\Psi}_{\cal L} = \left(
 \begin{array}{c}
 {\cal L}_{1}\\
 {\cal L}_{2}
\end{array}
\right),
\end{equation}
as well as the first and second generation 
right handed superfields,
\begin{equation}
\Psi_{\cal U} = \left(
 \begin{array}{c}
 {\cal U}_{1}\\
 {\cal U}_{2}
\end{array}
\right),
\Psi_{\cal D} = \left(
 \begin{array}{c}
 {\cal D}_{1}\\
 {\cal D}_{2}
\end{array}
\right), 
\Psi_{\cal E} = \left(
 \begin{array}{c}
 {\cal E}_{1}\\
 {\cal E}_{2}
\end{array}
\right),
\end{equation}
transform as covariant vectors under ${\rm U(2)}_{H}$,
We will denote them abbreviately by $\bar{\Psi}_{a}^{L}$ and $\Psi_{a}^{R}$.
We will introduce a set of supersymmetry breaking 
chiral superfields,
\begin{equation}
{\cal S}^{ab}, \quad {\cal A}^{ab}, \quad {\cal F}^{a} \quad (a,b=1,2),
\end{equation}
that transform under ${\rm U(2)}_{H}$ contravariantly as a symmetric tensor, an 
antisymmetric tensor and a vector respectively.
We will assume that only the auxiliary components of the 
flavor breaking superfields are non zero,
The most general form for the vevs of the flavor breaking fields is,
\begin{eqnarray}
<{\cal S}> &=& \left(
 \begin{array}{cc}
v_s   { e^{i \phi_{s}}} &  0 \\
 0 & { V}_{\cal S}  { e^{i \Phi_{S}}}
\end{array}
\right) \theta^{2}
\label{Svev}
, \\ 
<{\cal A}>&=& \left(
 \begin{array}{cc}
 0  &  v_{a}  { e^{i \phi_{a}}} \\
 - v_{a} { e^{i \phi_{a}}} &  0
\end{array}
\right) \theta^{2}, 
\label{Avev} \\ 
<{\cal F}>  &=& \left(
 \begin{array}{c}
 v_{f} e^{i \phi_{f}} \\
 { V}_{\cal F} e^{i \Phi_{F}}
\end{array}
\right) \theta^{2},
\label{Fvev}
\end{eqnarray}
where the $v_{s}$, $V_{\cal S}$, $v_{a}$, $v_{f}$ and $V_{\cal F}$ are real parameters.
We will assume the following particular hierarchy in the flavor breaking vevs,
\begin{equation}
\left( v_{f}, { v}_{a}, V_{S},  {V}_{\cal F}\right) =
 \left( \lambda^{2} , \lambda ^{2}, \lambda, 2 \lambda \right) M_{\rm F} \widetilde{m}.
 \label{vavf}
\end{equation}
We will also assume that
$v_s \ll { V}_{\cal S}$ and for practical purposes
set $v_{s}= 0$. 
Here $\lambda$ is a flavor breaking perturbation parameter,
$M_{\rm F}$ is the flavor breaking scale. We note that $\widetilde{m}$ is 
a new mass scale linked to the flavor violating susy-breaking fields.
We do not have yet a predictive model for the ${\rm U(2)}_{H}$ breaking, 
which is a relevant point under current investigation. 
The proposed vevs in Eq.~\ref{vavf}
are introduced ad-hoc. These ad-hoc assumptions will prove a posteriori 
to be very succesfull in reproducing fermion masses and mixings.
Furthermore, in the case that ${\rm U(2)}_{H} $ is a gauge symmetry broken spontaneously 
we expect the ${\rm U(2)}_{H}$-gauge fields to get masses of the order 
of the flavor breaking scale which can be very heavy in this scenario. 
Therefore any other phenomenological effects in the low energy 
model beyond the flavor structure it gives rise to in the soft supersymmetry
breaking sector would be very suppressed.

We will assume that the superpotential of the model is CP 
and ${\rm U(2)}_{H}$ symmetric. 
Therefore the only couplings allowed in the renormalizable superpotential
by the SM ${\rm SU(3)}_{C} \times {\rm SU(2)}_{L} \times {\rm U(1)}_{Y}$ vertical symmetry,
the ${\rm U(2)}_{H}$ horizontal symmetry and CP conservation 
are the third generation ones and the so called $\mu$--term, 
\begin{equation}
\lambda_{t} {\cal Q}_{3} {\cal U}_{3}{\cal H}_{u} +
\lambda_{b} {\cal Q}_{3} {\cal D}_{3}{\cal H}_{d} +
\lambda_{\tau} {\cal L}_{3} {\cal E}_{3}{\cal H}_{d} + \mu {\cal H}_{u} {\cal H}_{d}.
\label{superpo}
\end{equation}
We note that, in principle, two other couplings,
could be allowed in the superpotential:
${\cal L}_{3}{\cal H}_{u}$ and ${\cal Q}_{3}{\cal L}_{3}{\cal D}_{3}$. 
There are different ways to remove this unwanted couplings.
They could be forbidden imposing $R$--parity conservation
defined as $R = (-)^{3B + L +2S}$, where B is the barionic number, L the leptonic number
and S the spin. A third possibility would be to extend the ${\rm U(2)}_{H}$ symmetry
to the maximal ${\rm U(3)}_{H}$ horizontal symmetry. The breaking of the ${\rm U(3)}_{H}$ symmetry 
in the direction of the third generation would leave us with our ${\rm U(2)}_{H}$ symmetry,
in such a case this bilinear interaction would not be allowed by the
${\rm U(3)}_{H}$ symmetry. 
We also note that the couplings in the renormalizable superpotential cannot
carry complex phases since CP is an exact symmetry at this level.  
Therefore, at tree level the Yukawa matrices are generically of the form,
\begin{equation}
{\cal Y} = \left[
 \begin{array}{ccc}
0 & 0 & 0 \\
0 & 0 & 0 \\
 0 & 0 & y
\end{array}
\right], 
\label{Yukrad}
 \end{equation}
Additionally, trilinear soft supersymmetry 
breaking terms are generated by operators generically of the form,
\begin{eqnarray}
\sum_{{\cal Z}={\cal S}, {\cal A} } \frac{1}{M_{\rm F}}
\int d^{2} \theta   {\cal Z}^{ab}\bar{\Psi}^{L}_{a} \Psi_{b}^{R} {\cal H}_{\alpha}, \\
\frac{1}{M_{\rm F}} \int d^{2} \theta 
\left( {\cal F}^{a} \bar{\Psi}^{L}_{a}  \phi_{R}  +  \bar{\phi}^{L} 
{\cal F}^{a}  \Psi_{a}^{R}  \right) {\cal H}_{\alpha},
\label{softs}
\end{eqnarray}
where $M_{\rm F}$ is the flavor breaking scale,
$a=1,2$ are flavor indices, ${\cal H}_{\alpha} \, (~\alpha=u,d)$
represents any of the Higgs superfields. 
Soft supersymmetry breaking mass matrices can receive diagonal flavor conserving
contributions of the form, 
\begin{eqnarray}
 \frac{1}{M^{2}_{\rm F}}
\int  d^{4} \theta  
( \sum_{{\cal Z}, {\cal Z}^{\prime} ={\cal S},{\cal A} } 
  {\cal Z}^{\dagger} {\cal Z}^{\prime} +  {\cal F}^{\dagger}  {\cal F} ) 
  ( \Psi^{\dagger}  \Psi +  \phi^{\dagger}  \phi) . 
\label{softs}
\end{eqnarray}
Additionally non-diagonal flavor violating contributions arise from 
operators generically of the form,
\begin{eqnarray}
 \frac{1}{M^{2}_{\rm F}}
\int  d^{4} \theta  ( \sum_{{\cal Z}, {\cal Z}^{\prime} ={\cal S},{\cal A} } 
 \Psi^{\dagger a} {\cal Z}_{ac}^{\dagger}  {\cal Z}^{\prime cb} \Psi_{b} + 
 \Psi^{\dagger a} {\cal F}_{a}^{\dagger}  {\cal F}^{b} \Psi _{b} ), 
\label{softs}
\end{eqnarray}
where $a\neq b$.
Flavor violating supersymmetry-breaking fields cannot generate
masses for the gauginos. Therefore
we need to introduce at least one flavor--singlet chiral superfield, ${\cal G}$,
whose F-term component gets a non-zero vev
giving masses to gauginos from operators of the form,
\begin{equation}
\frac{1}{M}
\int d^{2} \theta 
{\cal G} 
\widetilde{\lambda} \widetilde{\lambda},
\label{gauginosmass}
\end{equation}
We note that $\left<{\cal G}\right>$ breaks 
supersymmetry but not the flavor symmetry. 
We will identify $M$ with the usual supersymmetry breaking messenger scale.
We note that the messenger scale is 
in general different from the flavor breaking scale even tough the flavor
breaking fields are supersymmetry breaking fields themselves.
The gaugino mass generated is given by 
$m_{\widetilde{\lambda}} = \left<{\cal G}\right>/M$.
The flavor singlet superfield responsible for generating gaugino masses, 
${\cal G}$, will couple to matter fields too generating soft trilinears couplings,
\begin{equation}
\frac{\kappa}{M}
\int d^{2} \theta 
{\cal G} \phi^{L} \phi^{R} {\cal H}_{\alpha},
\label{Gtril}
\end{equation}
where $\kappa$ is a real dimensionless coupling determined by some
unknown underlying renormalizable theory. For practical purposes we will
assume that $\kappa$ can take arbitrary values. 
Soft masses would also be generated by operators generically of the form,
\begin{eqnarray}
\frac{\eta}{M^{2}} 
\int  d^{4} \theta  {\cal G}^{\dagger} {\cal G}  
\left( \Psi^{\dagger} \Psi   + \phi^{\dagger} \phi \right) , \\
\frac{\eta^{\prime}}{M M_{F}} 
\int  d^{4} \theta  
{\cal G}^{\dagger} \phi^{\dagger} {\cal F}  \Psi.
\label{softs}
\end{eqnarray}
Here flavor indices have been omitted.
$\eta$ and $\eta^{\prime}$ are also real dimensionless 
couplings determined by the unknown underlying renormalizable theory.
Regarding the possible appearance of D-terms in the scalar potential. 
D-terms would appear when a local symmetry is spontaneously broken by
scalar fields,
which is not the case for the model under consideration. 
After the ${\rm U(2)}_{H}$ flavor and the CP symmetry are broken spontaneously
by the supresymmetry breaking fields defined in Eq.~\ref{Fvev}
the following boundary conditions for the soft trilinear matrices are generated
at the scale $M_{F}$,
\begin{equation}
{\bf A} = A \left[
 \begin{array}{ccc}
0 &  \sigma \lambda^{2} { e^{i \phi_{a}}}  & \sigma \lambda^{2} { e^{i \phi_{f}}}  \\
-  \sigma \lambda^{2} { e^{i \phi_{a}}} & \sigma \lambda { e^{i \Phi_{S}}}  
& 2 \sigma  \lambda { e^{i \phi_{F}}} \\
\sigma \lambda^{2}  { e^{i \phi_{f}}}  & 2 \sigma \lambda { e^{ i \phi_{F}}} & 1
\end{array}
\right], 
\label{Asoft}
 \end{equation}
where $A =   \kappa  m_{\widetilde{\lambda}} $ 
and the dimensionless parameter $\sigma$ is defined by
$\sigma = \widetilde{m} / {A}$. 
The mass parameter $\widetilde{m}$, defined in Eq.~\ref{vavf},
is a new mass scale introduced in the problem by the 
flavor violating susy-breaking fields.
We note that only one combination of the complex phases in Eq.~\ref{Asoft}
will be transmitted to the Yukawa matrices. For convenience one can remove
some of them from the soft trilinear matrix through a redefinition of the phases 
of the matter fields, even though they will appear in the soft mass matrices. 
Without any loss of generality 
we can adopt a flavor basis where the soft trilinear matrix takes the 
following form,
\begin{equation}
{\bf A} = A \left[
 \begin{array}{ccc}
0 &  \sigma \lambda^{2}  & \sigma \lambda^{2} { e^{- i \gamma}}  \\
-  \sigma \lambda^{2}  & \sigma \lambda & 2 \sigma  \lambda \\
\sigma \lambda^{2} { e^{-i \gamma}}  & 2 \sigma \lambda & e^{-i \phi}
\end{array}
\right], 
\label{AsoftD}
 \end{equation}
The phases $\gamma$ and $\phi$ are related with the phases in Eq.~\ref{Asoft}
by $\gamma = - (\phi_{f} + \Phi_{S} - \phi_{a} - \Phi_{F} )$
and $\phi=  (2 \Phi_{F}-\Phi_{S})$.
After the ${\rm U(2)}_{H}$ flavor breaking flavor violating 
soft mass matrices are also generated. In the flavor basis
adopted in Eq.~\ref{AsoftD} they take the following form, 
$$
\widetilde{{\cal M}}_{L,R}^{2} = \widetilde{m}^{2}_{f} \times 
$$
\begin{equation}
\left[
 \begin{array}{ccc}
1 + 5 \rho \lambda^{2} & \rho \lambda^{3}(2 e^{- i\gamma}-  e^{- i\phi^{\prime}})
& \rho^{\prime}  \lambda^{2} e^{-i(\gamma -\phi)}  \\
\rho \lambda^{3} (2 e^{i\gamma}-  e^{ i\phi^{\prime}})
 & 1  + 5 \rho \lambda^{2}  & 2 \rho^{\prime} \lambda e^{i\phi}  \\
\rho^{\prime}  \lambda^{2} e^{i(\gamma -\phi)} 
  & 2 \rho^{\prime} \lambda e^{-i\phi} & 1 + 5 \rho \lambda^{2}
\end{array}
\right],
\label{Msoft}
 \end{equation}
where, 
$\widetilde{m}^{2}_{f} =  \eta m^{2}_{\widetilde{\lambda}} $
and $\phi^{\prime}= (\phi+\phi_{a}-2\Phi_{F})$.
$\rho$ and $\rho^{\prime}$ are dimensionless parameters defined
by $\rho = \widetilde{m}^{2} / \widetilde{m}^{2}_{f}$ and 
$\rho^{\prime} = (\eta^{\prime}/\eta)
(\widetilde{m} / \widetilde{m}_{\tilde{\lambda}})$. 
We note that in this scenario the amount of flavor violation
in the soft mass matrices 
is determined not only by the powers of $\lambda$ in the
off-diagonal entries but also 
by the parameters $\sigma$, $\rho$ and $\rho^{\prime}$.
We note that in the limit $\tilde{m}\rightarrow 0$ all the flavor violation
will be suppressed. 
There are other interesting limits:
if $\eta^{\prime} \rightarrow 0$ the mixing between
third and first or second generation in the soft mass matrices is suppressed,
if $\tilde{m} \ll \tilde{m}_{f}$ the flavor mixing between
first and second generation in the soft mass matrices is also suppressed
and the sfermions masses will be nearly universal,
in the case $A\simeq \widetilde{m} \ll \tilde{m}_{f}$ only 
the contributions from soft masses to 
flavor violating processes would be suppressed while the
flavor violation in the soft trilinear matrices could be sizeable. 
\section{Radiative generation of Yukawa couplings \label{RadYuk}}
In the presence of flavor violation in the soft sector, the left and right handed components
of the sfermions mix. For instance, in the gauge basis 
the $6\times 6$ down--type squarks mass matrix is given by, 
\begin{equation}
{\cal M}^{2}_{D} = 
\left[
\begin{array}{cc}
\widetilde{{\cal M}}^{2}_{D_{L}} + v^{2}c^{2}_{\beta}{\cal Y}_{D}^{\dagger}{\cal Y}_{D}
 & ({\bf A}_{D}^{\dagger} c_{\beta} - 
\mu {\cal Y}_{D} s_{\beta}) v \\
 ({\bf A}_{D} c_{\beta} - 
 \mu {\cal Y_{D}^{\dagger}} s_{\beta})v & \widetilde{{\cal  M}}^{2}_{D_{R}} +
 v^{2}c^{2}_{\beta}{\cal  Y}_{D}{\cal Y}_{D}^{\dagger}
 \\
\end{array}
\right],
\end{equation}
where $\widetilde{\cal M}^{2}_{D_{R}}$ and $\widetilde{\cal M}^{2}_{D_{L}}$
are the $3 \times 3$ right handed and left handed soft mass matrices
(including D-terms), ${\bf A}_{D}$ is the $3 \times 3$ soft trilinear matrix,
${\cal Y}_{D}$ is the $3 \times 3$ tree-level Yukawa matrix.
$\tan \beta$ is the ratio of Higgs expectation values in the MSSM,
$\mu$ is the so-called mu-term and $v= s_{W} m_{W}/\sqrt{2\pi \alpha_{e}}=174.5$~GeV.
$\widetilde{\cal M}^{2}_{D}$ is diagonalized by a $6 \times 6$ unitary matrix, 
${\cal Z}^{D}$. 
The presence of flavor violating entries at tree level in the soft supresymmetry
breaking matrices will generate one loop contributions to the Yukawa matrices.
In general, the dominant finite 
one-loop contribution to the $3 \times 3$ down--type quark Yukawa matrix
including CP phases \cite{Ibrahim:2003ca}
is given by the gluino-squark loop,
\begin{equation}
({\cal Y}_{D})_{ab}^{\hbox{rad}} = \frac{\alpha_{s} }{3 \pi} 
m_{\widetilde{g}}^{*} 
\sum_{c} {\cal Z}^{D}_{ac} {\cal Z}_{(b+3)c}^{D*} 
B_{0}(m_{\widetilde{g}}, m_{\widetilde{d}_{c}}), 
\end{equation}
where $\widetilde{d}_{c}$ ($c =1, \cdot \cdot \cdot, 6$)
are mass eigenstates and $m_{\widetilde{g}}$ is the gluino mass.
$B_{0}$ is a known function that can be found elsewhere in the literature.
The radiatively corrected $3 \times 3$ down--type quark
mass matrix is given by,
\begin{equation}
{\bf m}_{D} = v c_{\beta} ( {\cal Y}_{D}+ {\cal Y}_{D}^{\hbox{rad}} ). \\
\end{equation}
We note that the effective supersymmetric model
proposed generates an approximately degenerate squark spectra.
In the squark degenerate case one obtains a simple expression
for ${\cal Y}_{D}^{\hbox{rad}}$, 
\begin{equation}
{\cal Y}_{D}^{\hbox{rad}} = \frac{2 \alpha_{s} }{3 \pi} 
m^{*}_{\widetilde{g}}  ({\bf A}_{D} - \mu {\cal Y}_{D} \tan {\beta}) 
F(m_{\widetilde{b}}, m_{\widetilde{b}},m_{\widetilde{g}}),
\end{equation}
where the function $F(x,y,z)$ is a form factor of the 
particles in the loop with units of [Mass]${}^{-2}$. $F(x,y,z)$ is
defined in Eq.~\ref{Ffun} of the appendix.
For the soft-trilinear texture in Eq.~\ref{Asoft} predicted by our model
one obtains a simple expression for 
the radiatively corrected down--type quark mass matrix,
\begin{equation}
{\bf m}_{D} =  \widehat{m}_{b}
\left[
 \begin{array}{ccc}
 0 & \omega \lambda^{2} &  \omega  \lambda^{2} e^{-i \gamma} \\
-\omega \lambda^{2} & \omega \lambda & 2 \omega \lambda \\
\omega  \lambda^{2} e^{-i \gamma} &  2 \omega \lambda & 1 
\end{array}
\right],
\label{mDmat}
\end{equation}
where $\omega$ encodes the dependence on the supersymmetric spectra.
For the case $m_{\widetilde{b}} \geq m_{\widetilde{g}}$, we obtain 
\begin{equation}
\omega = c_{\beta}\left(  \frac{v }{\widehat{m}_{b}}
\frac{2 \alpha_{s} }{3 \pi} \right) \left( \frac{m_{\widetilde{g}}}{m_{\widetilde{b}}} \right)
\left(\frac{\widetilde{m}}{m_{\widetilde{b}}}  \right).
\label{omegaEq}
\end{equation}
We emphasize that $\widetilde{m}$ is not any squark mass scale 
but a new mass scale introduced in the problem by the vevs of 
the flavor violating susy breaking fields, see Eqs.~\ref{Svev}-\ref{Fvev}.
The parameter $\widehat{m}_{b}$ defined as,
\begin{equation}
\widehat{m}_{b} = v c_{\beta} \left( y_{b} +  \omega_{b} (e^{-i\phi} - \frac{\mu}{A_{b}} y_{b} \tan \beta) \right), 
\label{mbottom}
\end{equation}
is approximately the running bottom mass.
$\omega_{b}= \omega \widehat{m}_{b}/ (vc_{\beta})$. 
The phase $e^{-i\phi}$ is  an overall phase absorbed in the definition of $\widehat{m}_{b}$
in Eq.~\ref{mbottom}, which has no observable implications.  
\subsection{Quark masses \label{Radmasses}}
The implications for fermion masses arising from a matrix similar to the
one in Eq.~\ref{mDmat} were studied in Refs.~\cite{Ferrandis:2004ri,Ferrandis:2004ng,Ferrandis:2004ti}.
In this subsection we briefly summarize results included in those references.
Although not diagonal in the gauge basis, the matrix ${\bf m}_{D}$ 
can be brought to diagonal form in the mass basis by a biunitary diagonalization,
$ ({\cal V}^{d}_{L})^{\dagger} {\bf m}_{D} {\cal V}^{d}_{R}
=  \left( m_{d}, m_{s}, m_{b} \right)$.
The down--type quark mass matrix given by Eq.~\ref{mDmat}
makes the following predictions for the quark mass ratios to leading order,
\begin{equation}
\frac{m_{d}}{m_{s}} = \lambda^{2} +{\cal O}(\lambda^{4}) , \quad 
\frac{m_{s}}{m_{b}} = \omega \lambda + {\cal O}(\lambda^{4})  . 
\label{downquarkrats}
 \end{equation}
We can relate $\lambda$ and $\omega$
with dimensionless fermion mass ratios. To first order,
\begin{equation}
\lambda = \left( \frac{m_{d}}{m_{s}}\right)^{1/2}, \quad 
\omega = \left(  \frac{m_{s}^{3}}{m_{b}^{2} m_{d}} \right)^{1/2}.
\label{lamomdown}
\end{equation}
Using these relations and the running quark masses
determined from experiment, see Ref.~\cite{Ferrandis:2004ti} for details,
we can determine $\lambda$ and $\omega$, they are given by 
$\lambda= 0.209 \pm 0.019$  and $\omega=  0.109 \pm 0.030$. 
We observe that constraints on the supersymmetric spectra can be derived from the
parameter $\omega$. To assess the viability of the model we must
check if it is possible for $\omega$ to reach the values required by 
the observed quark masses. 
From Eq.~\ref{omegaEq} we obtain 
the following unequality for $m_{\widetilde{g}} \leq m_{\widetilde{b}}$,
\begin{equation}
\omega = c_{\beta} 
\left(  \frac{v }{\widehat{m}_{b}}
\frac{2\alpha_{s} }{3\pi} \right) \left( \frac{\widetilde{m}}{m_{\widetilde{b}}} \right)
\left( \frac{m_{\widetilde{g}}}{m_{\widetilde{b}}} \right).
\label{omegaunEq1}
\end{equation}
Using the measured values $v=174.5$~GeV,
$m_{b}(m_{b})_{\rm MS}=4.2\pm0.2$ and
$\alpha_{s} = 0.117$ we obtain, 
\begin{equation}
\omega \leq 1.5 c_{\beta} \left( \frac{\widetilde{m}}{m_{\widetilde{b}}} \right). 
\label{omegaunEq2}
\end{equation}
Therefore the values of $\omega$ required by the measured quark masses 
can be easily reached without any ad-hoc 
tuning in the supersymmetric parameter space. Let us examine with some detail
the case $m_{\widetilde{g}}\approx m_{\widetilde{b}}$. For 
large $\tan\beta$, $\tan\beta = 50$, we obtain,
$$ 
\omega \approx 0.03 \left( \frac{\widetilde{m}}{m_{\widetilde{b}}} \right),
$$
which would require $\widetilde{m} \approx 3 m_{\widetilde{b}}$.  
On the other hand in the opposite gluino mass limit, 
$m_{\widetilde{g}} > 2 m_{\widetilde{b}}$, one obtains,
\begin{equation}
\omega = c_{\beta}\left(  \frac{v }{\widehat{m}_{b}}
\frac{2 \alpha_{s} }{3 \pi} \right) 
\left(\frac{\widetilde{m}}{m_{\widetilde{b}}}  \right)
\ln \left( \frac{{m}_{\widetilde{g}}}{m_{\widetilde{b}}}  \right).
\label{omegaEqlglu}
\end{equation}
For the large $\tan\beta$ case , $\tan\beta = 50$, 
we obtain similar results,
$$ 
\omega \approx 0.03 \left( \frac{\widetilde{m}}{m_{\widetilde{g}}} \right)
\ln \left(  \frac{m_{\widetilde{g}}}{m_{\widetilde{b}}}\right).
$$
If $\widetilde{m}\simeq m_{\widetilde{g}}$ we would need $m_{\widetilde{g}} \simeq 20 m_{\widetilde{b}}$.
This would imply $\widetilde{m} \simeq 20 m_{\widetilde{b}}$.
The mass matrix in Eq.~\ref{mDmat} is very succesful in reproducing the 
down-type quark mass ratios, but it 
cannot explain correctly the measured mass ratios in the 
up--type quark sector.  We will propose a simple solution. 
Let us assume that down and up type quark fields transform
differently under certain $Z_{2}$ symmetry. If this was the case
the fields ${\cal S}$, ${\cal A}$ and ${\cal F}$ would not generate
any mixing in the up type quark sector. Let us assume that there
is an extra $U(2)_{\rm H}$ symmetric tensor, ${\cal S}^{\prime}$, that gets a vev
of the form, 
\begin{equation}
<{\cal S}^{\prime}> =\left(
 \begin{array}{cc}
\lambda^{6}  &  0 \\
 0 & \lambda^{2}
\end{array}
\right) \theta^{2}.
\end{equation}
If the couplings of ${\cal S}^{\prime}$ with the up-type quark
fields are allowed by the $Z_{2}$ symmetry they would induce 
a soft trilinear matrix of the form,
\begin{equation}
{\bf A}_{U} = A_{t} \left[
 \begin{array}{ccc}
\sigma \lambda^{6} &  0 & 0 \\
0  & \sigma \lambda^{2} & 0 \\
0 & 0 & 1 
\end{array}
\right]. 
\label{AsoftU}
 \end{equation}
We note that we do not show any phases in the matrix $ {\bf A}_{U} $.
Possible phases in the diagonal entries of
$ {\bf A}_{U} $ are not physical to leading order since they can be absorbed 
through a redefinition of the phases of the matter fields.
Masses for the up and charm quarks are generated radiatively.
One can perform an analysis symilar to the analysis in the down-type
quark sector and obtain a simple expression for the 
radiatively corrected up--type quark mass matrix,
\begin{equation}
{\bf m}_{U} =  \widehat{m}_{t}
  \left[
 \begin{array}{ccc}
   \omega \lambda^{6} & 0 & 0 \\
  0  &  \omega \lambda^{2} &   0 \\
 0 &  0  &  1
\end{array}
\right],
\label{mUmat}
\end{equation}
where $\omega$ for $m_{\widetilde{t}} \geq m_{\widetilde{g}}$ 
is given in this case by, 
\begin{equation}
\omega \leq s_{\beta}\left(  \frac{v }{\widehat{m}_{t}}
\frac{2 \alpha_{s} }{3 \pi} \right) \left( \frac{m_{\widetilde{g}}}{m_{\widetilde{t}}} \right)
\left(\frac{\widetilde{m}}{m_{\widetilde{t}}}  \right),
\label{omegaEq}
\end{equation}
and $\widehat{m}_{t}$ is the normalized top quark mass given by,
\begin{equation}
\widehat{m}_{t} = v s_{\beta} \left( y_{t} +  \omega_{t} ( 1 - \frac{\mu}{A_{t}} y_{t} \cot \beta )\right), 
\end{equation}
with $\omega_{t}= \omega \widehat{m}_{t}/ (vs_{\beta})$. 
For the soft trilinear texture under consideration in Eq.~\ref{AsoftU}
one obtains the following predictions for the up--type quark mass ratios,
\begin{equation}
\frac{m_{u}}{m_{c}} = \lambda^{4} +{\cal O}(\lambda^{6}) , \quad 
\frac{m_{c}}{m_{t}} = \omega \lambda^{2} + {\cal O}(\lambda^{5}), 
\label{upquarkrats}
 \end{equation}
Again we can relate $\lambda$ and $\omega$
with dimensionless quark mass ratios, to first order,
\begin{equation}
\lambda = \left( \frac{m_{u}}{m_{c}}\right)^{1/4}, \quad 
\omega = \left(  \frac{m_{c}^{3}}{m_{t}^{2} m_{u}} \right)^{1/2}.
\label{lamomup}
\end{equation}
Using the running quark masses
determined from experiment, see Ref.~\cite{Ferrandis:2004ti} for details,
we obtain  $\lambda = 0.225 \pm 0.015$ and
$\omega =  0.071 \pm 0.018$. 
We note the similarity in the values 
for $\lambda$ and $\omega$ calculated in the up and down type quark sectors
from Eqs.~\ref{lamomdown} and \ref{lamomup}.
Let us examine with some detail
the case $m_{\widetilde{g}}\approx m_{\widetilde{t}}$. 
Using the measured top quark mass, $m_{t}=178$~GeV, we obtain the condition, 
$$
\omega \approx 0.02 s_{\beta}
 \left( \frac{\widetilde{m}}{m_{\widetilde{t}}} \right).
$$  
For instance, 
for large $\tan\beta$, $\tan\beta = 50$ we obtain 
$$
\omega \approx 0.02 
 \left( \frac{\widetilde{m}}{m_{\widetilde{t}}} \right).
$$ 
This constraint is compatible with the analogous
constraint arising from the down-type quark sector, which was 
$\omega \approx 0.03 ~(\widetilde{m}/m_{\widetilde{b}})$,
see Eq.~\ref{omegaunEq2}). Therefore no important splitting between the
sbottom and stop quark masses is required for the viability of the model. 
Indeed both constraints could be
satisfied simultaneously for $m_{\widetilde{b}}  \simeq 1.5 m_{\widetilde{t}}$, 
which is a non-trivial consistency check both of the model
and the ad-hoc vevs introduced in Eqs.~\ref{Svev}-\ref{Fvev}.
\subsection{Radiatively generated CP phase and CKM elements \label{radCKM}}
Finally, one can
calculate the CKM mixing matrix.
This is defined by ${\cal V}_{\rm CKM} =   
{\cal V}^{u\dagger}_{L} {\cal V}^{d}_{L}$.
We have seen in the previous section that in the simple model
here proposed the up-type quark mass matrix
is diagonal. Therefore the CKM matrix is given by
${\cal V}_{\rm CKM} =  {\cal V}^{d}_{L}$.
The diagonalization of the down-type quark mass matrix
in Eq.~\ref{mDmat} lead us to the following expression for
${\cal V}_{\rm CKM} $ to leading order in powers of $\lambda$,
\begin{equation}
\left[
 \begin{array}{ccc}
1 + 2i s_{\gamma} \lambda \omega -\lambda^{2}/2 & - \lambda (1 +  \omega \lambda (4 + 2i s_{\gamma})) & 
 \omega \lambda^{2} e^{-i \gamma}  \\
-\lambda (1 + 4\omega \lambda) &  \frac{1}{2} \left( \lambda^{2} +
4\omega^{2}\lambda^{2}
\right) -1
& 
2\omega \lambda
 \\
 \omega \lambda^{2} (2 - e^{ i \gamma}) & 2 \omega \lambda 
 & 1 - 2 \omega^{2} \lambda^{2}
\end{array}
\right]. 
\label{CKM}
\end{equation}
It is easy to check that this CKM matrix is unitary to order $\lambda^{3}$, i.e.
${\cal V}_{\rm CKM}^{\dagger} {\cal V}_{\rm CKM} = {\cal I} + {\cal O}(\lambda^{3})$.
We note that the model predicts that to leading order $\left| V_{us}\right| = \lambda$.
The measured value of $\left| V_{us}\right|$, $\left| V_{us}\right|_{\rm exp}=  0.220\pm 0.0026$,
agrees perfectly with the value of $\lambda$ as calculated from quark mass
ratios in Eqs.~\ref{lamomdown} and \ref{lamomup}. The model also predicts
that to leading order $\omega = \left| V_{cb}\right|/2\left| V_{us}\right|$.
Using the measured value for $\left| V_{cb}\right|$, $\left| V_{cb}\right|_{\rm exp}= 0.0413 \pm 0.0015$,
we obtain that $\omega = 0.093\pm0.005$. We note that this value of $\omega$ is 
surprisingly consistent with
the value calculated from quark mass 
ratios from Eqs.~\ref{lamomdown} and \ref{lamomup}. 
Finally using these values of $\lambda$
and $\omega$ we can predict $\left| V_{ub}\right|$ to be 
$\left| V_{ub}\right|=\omega  \lambda^{2} = 0.0045 \pm 0.0003$, which is consistent with the measured
value, $\left| V_{ub}\right|_{\rm exp}= 0.00367 \pm 0.00047$. 
Using again the previous values of $\lambda$ and $\omega$ calculated from 
 $\left| V_{us}\right|$ and  $\left| V_{cb}\right|$
we find the following predictions for $\left| V_{cs}\right|$ and $\left| V_{tb}\right|$,
$\left| V_{cs}\right|= 0.9795 \pm 0.0007 $ and $\left| V_{tb}\right|=0.9991\pm 0.0001 $, which are also
in agreement with experiment.
It is a trivial check to prove that
the angle $\gamma$ introduced in the parametrization of
the CKM matrix given in Eq.\ref{CKM} coincides 
with the standard definition for $\gamma=\phi_{2}$, 
\begin{equation}
\gamma = {\rm Arg} \left[ - \frac{V_{ud}V^{*}_{ub}} {V_{cd}V^{*}_{cb}} \right].
\end{equation}
The angle $\phi_{1}$ is defined as usual by,
\begin{equation}
\phi_{1} = {\rm Arg} \left[ - \frac{V_{cd}V^{*}_{cb}} {V_{td}V^{*}_{tb}} \right], 
\label{betadef}
\end{equation}
We note that we are using the notation $\phi_{1}$ and not the usual $\beta$ 
to avoid any confusion with the supersymmetric parameter $\tan\beta$, the ratio of the 
Higgs vevs in the MSSM.
The angle $\alpha=\phi_{3}$ can be obtained from,
$(\alpha + \phi_{1} + \gamma) = \pi$.
Using our parametrization for the CKM matrix 
we obtain to leading order in powers of $\lambda$ 
a simple relation between the angles $\phi_{1}$ and 
$\gamma$,  
\begin{equation}
\phi_{1} =  {\rm Arg} \left[ 2 - e^{-i \gamma}\right]
\label{betagamma}
\end{equation}
Using the measured value of the $\phi_{1}$ phase, 
$\phi_{1}^{\rm exp}= 23.3^{\circ }\pm 3.2^{\circ}~(2\sigma)$ 
we predict to leading order the phase $\gamma$
to be within the range $\gamma_{\rm theo} = 103^{\circ}\pm 13^{\circ}$.
This is far from the current $1\sigma$ global fit value for $\gamma$,
$\gamma_{\rm fit}= 61^{\circ} \pm 12^{\circ}$.
To find better agreement with the experimental constraints on $\gamma$
it is crucial to include in Eq.~\ref{betagamma} the next to leading order corrections
to the unitarity of the CKM matrix. We find that the only relevant correction
which affects the prediction for $\gamma$ is the correction to the 
element $V_{td}$, 
\begin{equation}
V_{td}^{\rm NLO} = (2 (1 + 5\omega \lambda) - e^{i\gamma} ) \omega \lambda^{2}.
\label{VtdNLO}
\end{equation}
Including this correction we predict the phase $\gamma$
to be within the range $\gamma_{\rm theo} = 91^{\circ}\pm 18^{\circ}$,
which intersects with the current $1\sigma$ global fit value for $\gamma$.
For this range of $\gamma$ $\left|V_{td}\right|$ is predicted to be,
$\left|V_{td}\right|=0.0043\pm 0.0005$. Finally $\left|V_{ud}\right|$
is predicted to be $\left|V_{ud}\right|= 0.9765\pm 0.0006$.
To sum up, in this model not only the first and second generation
fermion masses but also the CKM phase can be generated radiatively
in prefect agreement with current measurements.  
\section{Super CKM basis \label{SCKM}}
Overcoming the present experimental constraints on 
supersymmetric contributions to flavor changing and
CP violating processes is 
a necessary requirement for the consistency of any supersymmetric model 
\cite{susyFC}. 
In our scenario, as a consequence of the approximate radiative alignment 
between Yukawa and soft trilinear matrices there is
an extra suppression of the supersymmetric contributions to some of these processes.
Therefore for calculational purposes
it is convenient to rotate the squarks to the so-called superCKM basis
where this radiative alignment mechanism is manifest.

The superCKM basis is the 
basis where gaugino vertices are flavor diagonal ~\cite{masieroFCNC,Rosiek:1989rs,Misiak:1997ei}. 
In this basis, the entries in the soft trilinear matrices are directly proportional
to the corresponding contributions to flavor changing proceses. For instance,
the soft trilinear matrix ${\bf A_{D}}$
in the superCKM basis is given by,
\begin{equation}
 {\bf A}_{D}^{\rm{SCKM}} =
({\cal V}^{d}_{L})^{\dagger} {\bf A}_{D} {\cal V}^{d}_{R},
\end{equation}
where ${\cal V}^{d}_{L,R}$ are the down type quark diagonalization
matrices.
The soft trilinear matrix ${\bf A_{D}}$ is given by Eq.~\ref{AsoftD},
The Yukawa diagonalization matrices are given by
${\cal V}^{d}_{L}={\cal V}_{\rm CKM}$ in Eq.~\ref{CKM}
while ${\cal V}^{d}_{R}$ is completely determined to obtain real
mass eigenvalues after the diagonalization of ${\bf m}_{D}$.
We obtain, to leading order in $\lambda$,
\begin{equation}
{\rm Im}[{\bf A}_{D}^{\rm SCKM}] = (-) A_{b}  \left[
\begin{array}{ccc}
 0  &  2  \omega s_{\gamma} \sigma \lambda^{3} 
   & s_{\gamma} \sigma \lambda^{2}   \\
2 s_{\gamma} \sigma \omega \lambda^{3}
& 0 & 2 s_{\phi} \omega \lambda \\
s_{\gamma} \sigma \lambda^{2}  & 2 s_{\phi} \omega \lambda  & s_{\phi}
\end{array}
\right], 
\label{ASCKMIm}
\end{equation}
while the real part, ${\rm Re}[{\bf A}_{D}^{\rm SCKM}]$, is given by,
\begin{equation}
 A_{b}  \left[
\begin{array}{ccc}
\sigma \lambda^{3}  &  4 \sigma \omega \lambda^{3} (c_{\gamma}-1) &
\sigma \lambda^{2} ( c_{\gamma}-2) \\ 
4 \sigma \omega \lambda^{3} (c_{\gamma}+1) & \lambda \sigma &  2 \lambda (c_{\phi} \omega - \sigma) \\
\sigma \lambda^{2} ( c_{\gamma}+2)  & 2\lambda  (c_{\phi} \omega - \sigma) 
 & c_{\phi}+8\omega\sigma\lambda^{2}
\end{array}
\right], 
\end{equation}
We note that the entries $(2,1)$ and $(1,2)$ 
contain an additional suppression
factor $\omega\lambda$ compared with the soft trilinear matrix in the 
flavor basis, see Eq.~\ref{AsoftD}. This suppression is 
a consequence of the radiative alignment between Yukawa and soft trilinear matrices.
It is convenient when calculating supersymmetric contributions 
to flavor violating processes to use the parameters
$ (\delta_{ij}^{d})_{LR}$ defined as,
\begin{equation}
 (\delta_{ij}^{d})_{LR} = \frac{ v c_{\beta} ( {\bf A}_{D}^{\rm SKM})_{12}}{m^{2}_{\widetilde{q}}}.
\end{equation}
For consistency we also need to calculate the 
down-type squark soft mass matrices in the SCKM basis. For instance, 
for the left-handed soft mass matrix we obtain,
\begin{equation}
 (\widetilde{{\cal M}}^{2}_{D_{L}})^{\rm{SCKM}} =
({\cal V}^{d}_{L})^{\dagger}  (\widetilde{{\cal M}}^{2}_{D_{L}}) {\cal V}^{d}_{L},
\end{equation}
and analogously for the right handed soft mass matrix.
Assuming the soft trilinear texture from
Eq.~\ref{Msoft} we obtain for ${\rm Re}[( \widetilde{{\cal M}}^{2}_{D_{L}})^{\rm{SCKM}}]$,
to leading order in $\lambda$, 
\begin{equation}
m_{\widetilde{d}_{L}}^{2}  \left[
 \begin{array}{ccc}
1  &  y  \lambda^{3} 
  &  y^{\prime} \lambda^{2} \\
y \lambda^{3}  &  1 &  - 2 c_{\phi} \rho^{\prime} \lambda   \\
  y^{\prime} \lambda^{2}
&  -2 c_{\phi} \rho^{\prime} \lambda  &  
(1 + (  8c_{\phi}\rho^{\prime} \omega -  5\sigma \lambda^{2} )\lambda^{2})
\end{array}
\right].
\end{equation}
Here $m_{\widetilde{d}_{L}}^{2}$, $y$ and $y^{\prime}$ are defined as,
\begin{eqnarray}
m_{\widetilde{d}_{L}}^{2} &=& m_{\widetilde{f}}^{2}  ( 1 + ( 1 +  5\rho )\lambda^{2}) ,\\
y &=& ( 4  \rho^{\prime} \omega ( c_{\gamma} c_{\phi} - 2 c_{\phi}) +
 \rho (c_{2\phi^{\prime}} - 2 c_{\gamma})) ,\\
y^{\prime} &=&  (c_{(\phi-\gamma)} - 2 c_{\phi}) \rho^{\prime}.
\end{eqnarray}
We note that if the gluino mass is of the same order than the squark masses,
$m_{\tilde{\lambda}} \simeq \widetilde{m}_{f}$,
$\widetilde{m}_{f} \simeq \widetilde{m}$
and $\eta^{\prime}  \simeq \eta$ we expect that $\rho \simeq \rho^{\prime}$.
In that case the coefficient 
$y$ simplifies to $y \approx   \rho (c_{2\phi^{\prime}} - 2 c_{\gamma}) $
since $\omega \approx 2 \lambda^{2}$.
Furthermore if $\rho^{\prime}/  \rho = (\eta^{\prime}/\eta) (\widetilde{m}_{f}^{2}/\widetilde{m} m_{\tilde{\lambda}})$
the limit  $\rho^{\prime} \gg \rho$ would correspond to $\eta^{\prime}\gg
\eta$ or $\widetilde{m}\ll \widetilde{m}_{f}^{2}$ or $m_{\tilde{\lambda}} \ll \widetilde{m}_{f}^{2}$,
if this is not the case we would expect the $\rho$ term to dominate. 
The imaginary component, ${\rm Im}[({\cal M}^{2}_{D_{L}})^{\rm{SCKM}}]$,
is given to leading order in $\lambda$ by, 
\begin{equation}
m_{\widetilde{d}_{L}}^{2}  \left[
 \begin{array}{ccc}
0 &  -z \lambda^{3} 
&  - z^{\prime} \lambda^{2} \\
z \lambda^{3} 
 &  0 &  - 2 s_{\phi} \rho^{\prime} \lambda   \\
z^{\prime} \lambda^{2}
&  2 s_{\phi} \rho^{\prime} \lambda  &  0 \\
\end{array}
\right].
\end{equation}
where, 
\begin{eqnarray}
z &=& ( 4\rho^{\prime} \omega c_{\gamma}s_{\phi}  +
 \rho (s_{2\phi^{\prime}} - 2 s_{\gamma})) ,\\
 z^{\prime} &=& (2 s_{\phi} - s_{(\phi-\gamma)} ) \rho^{\prime}.
  \end{eqnarray}
If $\rho \simeq \rho^{\prime}$ the coefficient 
$z$ reduces to $z \approx   \rho (s_{2\phi^{\prime}} - 2 s_{\gamma}) $
since $\omega \approx 2 \lambda^{2}$.
Assuming the soft trilinear texture from
Eq.~\ref{Msoft} we obtain for ${\rm Re}[({\cal M}^{2}_{D_{R}})^{\rm{SCKM}}]$,
to leading order in $\lambda$, 
\begin{equation}
m_{\widetilde{d}_{R}}^{2}  \left[
 \begin{array}{ccc}
1  &   r \lambda^{3} 
  &  r^{\prime} \lambda^{2} \\
r  \lambda^{3}  &  1 &  - 2 c_{\phi} \rho^{\prime} \lambda   \\
 r^{\prime}  \lambda^{2}
&  -2 c_{\phi} \rho^{\prime} \lambda  &  
(1 + (  8c_{\phi}\rho^{\prime} \omega -  5\sigma \lambda^{2} )\lambda^{2})
\end{array}
\right].
\end{equation}
Here, 
\begin{eqnarray}
m_{\widetilde{d}_{R}}^{2} &=& m_{\widetilde{f}}^{2}  ( 1 + ( 1 +  5\rho )\lambda^{2}), \\
r &=& ( 4  \rho^{\prime} \omega (  c_{(\gamma-\phi)} + 2 c_{\phi}) +
\rho (c_{2\phi^{\prime}} - 2 c_{\gamma})), \\
r^{\prime} &=&  (c_{(\gamma-\phi)} + 2 c_{\phi}) \rho^{\prime}. 
\end{eqnarray}
Again if $\rho \simeq \rho^{\prime}$ the coefficient 
$r$ reduces to $r \approx  \rho (c_{2\phi^{\prime}} - 2 c_{\gamma}) $
since $\omega \approx 2 \lambda^{2}$.
The imaginary component, ${\rm Im}[({\cal M}^{2}_{D_{R}})^{\rm{SCKM}}]$,
is given to leading order in $\lambda$ by, 
\begin{equation}
m_{\widetilde{d}_{R}}^{2}  \left[
 \begin{array}{ccc}
0 &  - t \lambda^{3}  
&   - t^{\prime} \lambda^{2} \\
 t \lambda^{3} 
 &  0 &  - 2 s_{\phi} \rho^{\prime} \lambda   \\
  t^{\prime} \lambda^{2}
&  2 s_{\phi} \rho^{\prime} \lambda  &  0 \\
\end{array}
\right],
\end{equation}
where, 
\begin{eqnarray}
t &=&  (s_{2\phi^{\prime}} - 2 s_{\gamma} ) \rho , \\
t^{\prime} &=&  (s_{(\gamma-\phi)} - 2 s_{\phi}  ) \rho^{\prime}.
\end{eqnarray}
It is also convenient when calculating supersymmetric contributions 
to flavor violating processes to use
the couplings $ (\delta_{ij}^{d})_{LL}$ defined by,
\begin{equation}
 (\delta_{ij}^{d})_{LL} = \frac{ ({\widetilde{\cal M}}_{D_{L}}^{2})^{\rm SCKM}_{ij}}{m^{2}_{\widetilde{d}_{L}}}.
\end{equation}
One can define 
analogously the 
$ (\delta_{ij}^{d})_{RR}$ couplings.

\section{Suppressed contributions to EDMs \label{EDMs}}
In an unconstrained minimal supersymmetric
standard model (MSSM) the generic contribution to the neutron EDM
\cite{Ellis:1982tk,wyler1,Polchinski:1983zd}
is around eights orders of magnitude larger than the SM contribution,
{\it i.e.} about four orders of magnitude above the current experimental constraint
\cite{EDMscomment}.
This is the so called SUSY CP problem or to be 
more especific the flavor conserving SUSY CP problem.
The disparity between the current
experimental constraint and the generic supersymmetric contribution in the MSSM 
is due to the, in principle, allowed presence of CP phases in the superpotential
and in the soft supersymmetry breaking sector. 
Numerous papers have examined this topic in the context
of supersymmetric models \cite{EDMs}
and a few solutions have been 
proposed, which were summarized in the introduction.
We will explain with some detail how generic supersymmetric
models for radiative mass generation can ameliorate this problem.
We will analyze separately the one-loop, 
two-loop and higher order contributions to EDMs.

Interestingly, the one loop supersymmetric
contributions to EDMs always appear as combinations 
of six possible physical phases of the generic form
\cite{Dugan:1984qf,Dimopoulos:1995kn,Ibrahim:1998je},
\begin{equation}
{\rm Arg}(A^{*}m_{\widetilde{\lambda}}), \quad {\rm Arg}(B^{*}\mu m_{\widetilde{\lambda}}),
\label{EDMphases}
\end{equation}
where $A$ are first generation flavor diagonal trilinear soft supersymmetry breaking parameters,
$m$ are gaugino masses, $B$ is the bilinear soft supersymmetry breaking term
and $\mu$ is the superpotential bilinear term. In the special case of
universal soft supersymmetry breaking terms these 
reduce to only two physical phases.
First let us focus our attention on the term
${\rm Arg}(B^{*}\mu m_{\widetilde{\lambda}})$.
The tree level gaugino masses are flavor conserving
parameters generated by the supersymmetry breaking 
flavor singlet field ${\cal G}$ as we explained before
and as a consequence cannot carry complex phases,
which are linked to flavor breaking vevs. 
The $\mu$ term is allowed in the CP conserving
superpotential at tree level. This parameter is linked to the flavor blind
operator ${\cal H}_{u}{\cal H}_{d}$, which obviously 
cannot carry CP phases at tree level. 
For the same reason
the bilinear soft supersymmetry breaking term, $B$,
is also a real parameter since the term ${h}_{u}{h}_{d}$
in the scalar potential is also a flavor singlet. 
Regarding the contributions of the form ${\rm Arg}(A^{*}m_{\widetilde{\lambda}})$.
In the case of the neutron and mercury EDMs the relevant
terms arise from the up and down quarks EDMs which are
${\rm Arg}(A^{*}_{u}m_{\widetilde{\lambda}})$ and
${\rm Arg}(A^{*}_{d}m_{\widetilde{\lambda}})$. We have seen that in our model
the $3\times3$ matrix ${\bf A}_{U}$ is diagonal, see Eq.~\ref{AsoftU}.
The diagonal entries carry no CP phases. Even whether they existed 
they could be absorbed in a redefinition of the phases of the up-type matter fields.
Furthermore, the entry (11) of the $3\times3$ soft trilinear matrix in the down-type squark sector
corresponding to $A_{d}$ in Eq.~\ref{ASCKMIm},
is in general real. 
Therefore all the one-loop contributions
to the EDMs in this model are exactly zero. 

Regarding the two loop contributions to EDMs. 
It has been pointed out that the two-loop supersymmetic contributions 
can also constraint the supersymmetric parameter space, even though 
not so severly as the one-loop contributions \cite{Chang:1998uc,Pilaftsis:1999td,Pilaftsis:2002fe}. 
For instance, two-loop contributions of the Barr-zee type to $d_{d}$
exist with an exchange of stops and the CP-odd Higgs. This contribution is
of the form \cite{Chang:1998uc}, 
\begin{equation}
\left( \frac{d_{d}}{e} \right)^{\widetilde{t}}_{\rm 2-loop} =
\frac{-\alpha_{e}}{77 \pi^{3}} \frac{m_{d}m_{t}}{v^{2}} \frac{ \mu \sin(2 \theta_{t} )}{s_{2\beta} m^{2}_{A}}
\sin \delta_{t} C ( \widetilde{t}_{1} ,  \widetilde{t}_{2} , A)
\label{2loopEDM}
\end{equation}
where $\delta_{t} = {\rm Arg}[A_{t} + \cot \beta \mu^{*}]$, $\theta_{t}$ is the
stop mixing, 
$v=175$~GeV, $A$ is the CP-odd Higgs and $C$ stands for a dimensionless
two-loop form factor which can be found in Ref.~\cite{Chang:1998uc}. 
For the large $\tan\beta$ case under consideration $\delta_{t} \approx {\rm Arg}[A_{t}] = 0$
since $A_{t}$ and $\mu$ are real parameters. On the other hand there is an analogous 
contribution from bottom squarks that is proportional 
to ${\rm Im}[A_{b}e^{i\delta_{b}}]$ with
$\delta_{b}= {\rm Arg}[A_{b} + \tan \beta \mu^{*}]$ For the large $\tan\beta$ 
case under consideration the second term in $\delta_{b}$ would dominate 
in general and we obtain that the contribution is proportional to
$- A_{b} \sin \phi$, as can be seen from Eq.~\ref{ASCKMIm}.
We obtain that for CP-odd Higgs masses above $1$~TeV the resulting contribution
to the neutron EDM is below the current experimental constraint, $|d_{n}| < 10^{-25}$~e.cm.

One may wonder if the previous arguments for the cancellation
of the one loop contribution to EDMs could be extended in a variant of
this model to not only suppress but cancel the two-loop contributions.
We note that if the flavor model generates hermitian soft trilinear 
matrices the two-loop contributions, which are 
proportional to the phases of $A_{b}$ and $A_{t}$, would be zero.
It may be possible in principle that small phases are
generated in the ``flavor conserving'' parameters in the lagrangian, as the
$\mu$-term, the $B$ parameter or the gaugino masses.
Nonetheless, we note that many of the higher order operators which 
could contribute to the radiative corrections
to the $\mu$ term have to be flavor conserving operators of the form:
$\propto {\cal Z}^{\dagger ab}{\cal Z}_{ab}$ (${\cal Z}={\cal S},{\cal A}$) or
$\propto {\cal F}^{\dagger a}{\cal F}_{a}$, etc,$\cdot\cdot\cdot$. 
After the breaking of the flavor symmetry
these operators cannot generate complex phases since the presence 
of a complex phase would be an indication of flavor violation. 
To sum up, because of the intrinsic flavor 
off-diagonal nature of the CP violating phases in this model, the one loop
contributions to EDMs are zero and two and higher order contributions
are suppressed below experimental limits. 
\section{Contributions to direct and indirect CP violation  
in the Kaon system \label{Ksystem}} 
\subsection{$\epsilon$ \label{epsilon}}
The measure of indirect CP violation in the Kaon system is given by the
parameter $\epsilon$ defined by,
\begin{equation}
\epsilon  = \frac{A(K_{L}\rightarrow \pi\pi)}{A(K_{S}\rightarrow \pi\pi)}
 \approx \frac{e^{i\pi/4}}{\sqrt{2} \Delta m_{K} }{\rm Im} \left[{\cal M}_{12}  \right]
\label{epsK}
\end{equation}
where $\Delta m_{K}$ is the
$K_{L}K_{S}$ mass difference,
${\cal M}_{12}= {\cal M}(K^{0}) =  \left< K^{0} \left| {\cal H}_{\rm eff}^{\Delta S=2} \right| \bar{K}^{0} \right> $
is the $K^{0}\bar{K}^{0}$ mixing amplitude 
and ${\cal H}_{\rm eff}^{\Delta S=2}$ is
the effective $\Delta S=2$ hamiltonian.
The parameters $\Delta m_{K}$ and $\left|\epsilon\right|$ 
have received considerable attention in supersymmetric models
since their measured values have been known with good precision
for long time \cite{masieroFCNC,eKsusy,Ciuchini:1998ix}.
We will separate the SM and supersymmetric contributions to the
mixing amplitude in the form ${\cal M}_{12} = {\cal M}_{12}^{\rm SUSY}
+{\cal  M}_{12}^{\rm SM}$. 
We define $\phi_{\epsilon}$ and $\phi_{\epsilon}^{\prime}$ as the phases of
the SM and the supersymmetric contributions respectively, 
${\cal M}_{12}^{\rm SUSY} = | {\cal M}_{12}^{\rm SUSY} | e^{i\phi_{\epsilon}^{\prime}}$
and ${\cal M}_{12}^{\rm SM} = |{\cal  M}_{12}^{\rm SM} | e^{i\phi_{\epsilon}}$.
It is also convenient to introduce the ratio $R_{K} =  | {\cal M}_{12}^{\rm SUSY}| 
/ |{\cal M}_{12}^{\rm SM}|$. 
This ratio and the complex phase $\phi_{\epsilon}^{\prime}$ are constrained
by the experimental measurements of $\Delta m_{K}$ and $\left|\epsilon\right|$.
We can expand in powers of $R_{K}$ to obtain the following expression for
the new physics contributions to $\Delta m_{K}$ and $\left|\epsilon\right|$,
\begin{eqnarray}
\frac{ \Delta m_{K} - \Delta m_{K}^{\rm SM} }{\Delta m^{\rm SM}_{K}}
&\approx& 
 R_{K}  \frac{\cos(2 \phi_{\epsilon}^{\prime})}{\cos(2 \phi_{\epsilon})}
\label{dmKapr} \\
\frac{ \left|\epsilon\right| - \left|\epsilon^{\rm SM}\right|}{\left|\epsilon^{\rm SM}\right|}
&\approx&  
R_{K}  \frac{\sin(2 \phi_{\epsilon}^{\prime})}{\sin(2 \phi_{\epsilon})}
\label{eKapr}
\end{eqnarray}
We note that
$\Delta m_{K}$ has been measured with an uncertainity of approximately $0.2\%$,
$\Delta m_{K}^{\rm exp} = (3.490 \pm 0.006)\times 10^{-12}$~MeV.
In the SM roughly $70\%$ of the measured $\Delta m_{K}$
is described by the real parts of the box diagrams with charm quark and top quark exchanges.
Some non-negligible contribution comes from the box diagrams with 
simultaneous charm and top exchanges while approximately the remaining $20\%$ 
of the measured $\Delta m_{K}$ is attributed to long distance contributions.
On the other hand these are potentially sizeable and up to date incalculable.
While a precise prediction is not possible, the observation
is roughly compatible with the SM expectation. 
Assuming 
that the supersymmetric contribution saturates one half of 
the experimental measurement
and using Eq.~\ref{dmKapr}
we can obtain an approximate upper constraint on $R_{K}$, 
\begin{equation}
R_{K} \cos(2 \phi_{\epsilon}^{\prime})
\lesssim  \frac{1}{2} \cos(2 \phi_{\epsilon}) .
\label{dmkcons} 
\end{equation}
$\left|\epsilon\right|$ has been measured with an uncertainity of approximately $0.6$\%. 
A fit to the $K\rightarrow \pi \pi$ data yields
$\left|\epsilon\right|= (2.284 \pm 0.014) \times 10^{-3}$.
The calculation of $|\epsilon|$ is also affected by 
large distance corrections. 
It is usual when calculating $|\epsilon|$
to input the experimental 
measurement for $\Delta m_{K}$ in the denominator of Eq.~\ref{epsK}.
We will allow again for the supersymmetric contribution to saturate one half of 
the experimental measurement.
The resulting constraint on $R_{K}$ and $\phi_{\epsilon}^{\prime}$
can be expressed in the form,
\begin{equation}
R_{K} \sin(2 \phi_{\epsilon}^{\prime})
\lesssim  
\frac{1}{2} \sin(2 \phi_{\epsilon}^{\prime}).
\label{eKcons}
\end{equation}
Using the usual Wolfenstein parametrization of the CKM matrix to second order 
in powers of $\lambda$
we obtain that the dominant SM contribution to ${\cal M}_{12}$ is proportional to $(V_{cs} V^{*}_{cd})^{2}$. 
This is given by $V_{cs} V^{*}_{cd} = - 
\lambda (1-\frac{\lambda^{2}}{2}) (1 + A^{2} \lambda^{4} (\rho - i \eta))$.
This can be written as $V_{cs} V^{*}_{cd} 
\approx - \lambda (1-\frac{\lambda^{2}}{2}) e^{- i \phi_{\epsilon}}$ 
where $\phi_{\epsilon}$ is defined as 
$\phi_{\epsilon} = {\rm tan}^{-1} (\eta A^{2} \lambda^{4})$. 
This is a very small number.
Using updated extractions of $A$, $\eta$ and $\lambda$ we obtain 
$\phi_{\epsilon} \approx 0.03^{\circ}$. Therefore the 
constraints from Eqs.~\ref{dmkcons} and \ref{eKcons} can be written in the form, 
\begin{eqnarray}
R_{K} \cos(2 \phi_{\epsilon}^{\prime}) &\lesssim  &0.5, 
\label{cons1} \\
R_{K} \sin(2 \phi_{\epsilon}^{\prime}) &\lesssim & 6 \times 10^{-4}.
\label{cons2}
\end{eqnarray}
Next we need to calculate the expressions for $R_{K}$ and $\phi_{\epsilon}^{\prime}$ in our model.
We will see that $\phi_{\epsilon}^{\prime}$ in the model under consideration is completely
determined given the measured value of $\gamma$ and the quark mass ratios.
Therefore the previous constraints in Eq.~\ref{cons1} and \ref{cons2} will
translate into lower bounds on the squark mass spectra. The supersymmetric 
contribution to ${\cal M}_{12}$ contains contributions proportional to the
different $\delta$ couplings in the soft supersymmetry breaking sector. 
There are qualitatively four different contributions to  ${\cal M}_{12}^{\rm SUSY}$
that in our model, to leading order in powers of $\lambda$, 
take the following values, 
\begin{eqnarray}
 (\delta_{LR}^{d})^{2}_{12}+(\delta_{RL}^{d})^{2}_{12}  &=& 2 \delta^{12}_{\rm LR}
( \frac{5}{4} c_{2\gamma} + \frac{11}{4}  - is_{2\gamma}),
    \\
 (\delta_{LR}^{d})_{12} (\delta_{RL}^{d})_{12} &=& \delta^{12}_{\rm LR}
 ( \frac{5}{4} c_{2\gamma}-
\frac{5}{4} - is_{2\gamma}),
\label{deltasLR12}
 \end{eqnarray}
where,
\begin{equation}
\delta_{\rm LR}^{12} = 8 \omega^{2} \lambda^{6}  c^{2}_{\beta}
 \frac{ v^{2}}{m^{2}_{\widetilde{b}}}  \frac{\widetilde{m}^{2}}{m^{2}_{\widetilde{b}}}, 
\end{equation}
and,
\begin{eqnarray}
  (\delta_{LL}^{d})^{2}_{12}+(\delta_{RR}^{d})^{2}_{12} &=& 
 2 \delta_{\rm LL }^{12} (e^{-2i \phi^{\prime}} - 2 e^{-i\gamma})^{2} ,
  \\
(\delta_{LL}^{d})_{12}(\delta_{RR}^{d})_{12} &=& \delta_{\rm LL}^{12}
  (e^{-2i\phi^{\prime} }- 2e^{-i\gamma})^{2},
\label{deltasLL12}
\end{eqnarray}
where,
\begin{equation}
\delta^{12}_{\rm LL} = \rho^{2} \lambda^{6} .  
\label{deltaLLRR12}
\end{equation}
For instance, we noted in Sec.~\ref{Radmasses} that 
for large $\tan\beta$ and $m_{\widetilde{g}} \approx m_{\widetilde{b}}$ 
the parameter $\widetilde{m}$ (new mass scale introduced by
the flavor violating susy breaking fields) is required to be approximately
$\widetilde{m} \approx 3 m_{\widetilde{b}}$. Therefore the
LR couplings in our model have an important additional suppression
factor, $\delta_{LR}^{2}/\delta^{2}_{LL} \propto c_{\beta}^{2} \omega^{2} v^{2}/m^{2}_{\widetilde{b}}$. 
As a consequence in this model the $\delta_{LL}$ and $\delta_{RR}$ 
couplings dominate the contribution to ${\cal M}_{12}^{\rm SUSY}$.
All the $\delta_{LL}$ and $\delta_{RR}$ 
contributions to the Wilson coefficients 
can be added up in a simple expression. We obtain,
\begin{eqnarray}
 {\cal M}_{12}^{\rm SUSY}&=&
\frac{\alpha^{2}_{s} m_{K} f^{2}_{K} X^{2}_{K}}{60 m^{2}_{\widetilde{b}}}
\left[
 \left( (\delta_{LL}^{d})^{2}_{12}+(\delta_{RR}^{d})^{2}_{12} \right) \right. 
 \nonumber \\
 && \left. \times D(x) - (\delta_{LL}^{d})_{12} (\delta_{RR}^{d})_{12} C(x) \right].
 \label{M12KLR}
 \end{eqnarray}
Here $X^{2}_{K}$ is a dimensionless factor defined as 
$X_{K}^{2}= m^{2}_{K}/(m_{s} +m_{d})^{2}$,
numerically $X_{K}\approx 4.07$, $f_{K}$ is 
the K-meson decay constant and 
$x$ is the gluino-squark mass ratio squared,
$x= m^{2}_{\tilde{g}}/m^{2}_{\widetilde{b}}$.
$C(x)$ and $D(x)$ are given by
$C(x)=C_{f} f(x)+ C_{g}g(x)$ and $D(x)=D h(x)$ where 
$f(x)$, $g(x)$ and $h(x)$ are dimensionless form factors defined in 
Eqs.~\ref{fx}, \ref{gx} and \ref{hx} of the appendix.
The functions $f(x)$, $g(x)$ and $h(x)$ have been conveniently normalized so that 
in the limit $m_{\widetilde{g}}\approx m_{\widetilde{b}}$ they tend to 1. 
We note that in our approach the constant 
coefficients $C_{f}$, $C_{g}$ and $D$ 
absorb the dependency on the method of calculation of the hadronic matrix elements
as well as the renormalization effects on the Wilson 
coefficients. 
Following Ref.~\cite{Ciuchini:1998ix}, where lattice 
QCD methods were used to calculate the relevant
hadronic matrix elements and including NLO renormalization effects to the 
Wilson coefficients,
we obtain the following numerical values, 
$C_{f} \approx 9.13$, $C_{g}\approx 0.75$
and $D\approx 0.002$.
We note that the naive vacuum insertion approximation at tree level
gives the values $C_{f}   \approx 1.8$, 
$C_{g}\approx 0.067$ and
$D \approx 0.0055$,
which are significantly different. 
We also note that these constant parameters, $C_{f}$,$C_{g}$ and $D$, do not depend on the 
flavor mixing structure in the soft supersymmetric breaking sector.
From the numerical values of these coefficients we 
note that the contribution of the form $\delta_{LL} \delta_{RR}$ 
in Eq.~\ref{deltasLL12} dominates the
supersymmetric contribution.
Finally, using the well known expression for the SM contribution to
the amplitude, see for instance Eq.(3.39) in Ref.~\cite{Buras:2001pn},
we can write the ratio of the supersymmetric contribution 
over the SM amplitude in the form,
\begin{equation}
R_{K} 
\approx 
2  \frac{ \eta^{2}_{K} }{ m^{2}_{\widetilde{b}}} f(x)
 \rho^{2} \lambda^{6} 
 ( \frac{5}{4} - \cos(\gamma-2\phi^{\prime}))^{1/2}
 \label{RK}
   ,\end{equation}
where,
\begin{equation}
\eta_{K} \approx
 \frac{\alpha_{s} \pi X_{K} C_{f}^{1/2}}{
\sqrt{5} G_{F } V_{cs} V_{cd} B_{K}^{1/2} \eta_{1}^{1/2} m_{c}} \approx 666~\hbox{TeV}.
\end{equation}
Here $\hat{B}_{K}= 0.85 \pm 0.15$ is a renormalization group invariant
form of the B parameter arising from the hadronic matrix element, 
$m_{c}$ is the charm quark mass and $\eta_{1}$ is a short distance
QCD correction factor, at NLO $\eta_{1}$ is given by $\eta_{1}=1.38 \pm 0.20$
\cite{Buras:2001pn}.  
Using for $\lambda$  the value determined from 
CKM elements, $\lambda\approx 0.22$, and for $\gamma$ the $1\sigma$ global fit
$\gamma_{\rm fit} = 61^{\circ}\pm 11^{\circ}$ we obtain, 
\begin{equation}
R_{K} 
\approx
\left( \frac{\widetilde{m}}{m_{\widetilde{b}}} \right)^{4}
\left( \frac{10 ~\rm TeV }{ m_{\widetilde{b}}}\right) ^{2} h(x) ,
\end{equation}
The phase of the supersymmetric contribution, $\phi_{\epsilon}^{\prime}$,
can be calculated from Eq.~\ref{deltasLL12}.
This phase depends strongly on the phase $\phi^{\prime}$, which
is not constrained by the CKM matrix. For instance, for $\phi^{\prime}=0$
we obtain,
\begin{equation}
\tan (\phi_{\epsilon}^{\prime})
\approx  
\frac{s_{\gamma}( 1 - 2 c_{\gamma}) }{( \frac{1}{4} -  c_{\gamma} - c_{2\gamma} )}.
 \end{equation}
For $\gamma=60^{\circ}$ we would obtain $\phi^{\prime}_{\epsilon}=0$. 
If this was the case the measurement of $\left|\epsilon\right|$
would not constraint the supersymmetric spectra, see 
Eq.~\ref{cons2}. 
The only phase independent constraint comes from the
$\Delta m_{K}$ measurement, see Eq.~\ref{cons1}.
Let us analyze with more detail
the large $\tan\beta$ case. For $\tan\beta=50$ 
and $m_{\widetilde{g}} \approx m_{\widetilde{b}}$
we noted in Sec.~\ref{Radmasses}
that for the hierarchy of flavor breaking vevs
postulated in Eqs.~\ref{Svev}-\ref{Fvev} we need to have
$\widetilde{m} \approx 3 m_{\widetilde{b}}$, {\it i.e.}
$\rho \approx 9$. Therefore using Eqs.~\ref{cons1},\ref{cons2} and
\ref{RK} we obtain for $\tan\beta=50$ the constraints, 
\begin{eqnarray}
m_{\widetilde{b}} &\gtrsim& 130~{\rm TeV}, ~~~\Delta m_{K}, \\
m_{\widetilde{b}} &\gtrsim& 3670~{\rm TeV},~~~~\left| \epsilon \right| ~(\phi^{\prime} \neq 0).
\end{eqnarray}
To obtain the second constraint we assume that
the phase $\phi^{\prime}$ is arbitrary. If this phase was zero
as we mentioned this second constraint 
would not be effective. 
We would like to emphasize that these constraints
are non-generic. They only apply for the case large $\tan\beta$ case study
and the particular texture considered in this paper. 
This case study must be considered a worst case scenario
for this kind of models. Several variants of this model may allow us to
lower considerably these bounds. For instance we could lower the bounds
at the price of less predictivity in the fermion sector by increasing
the number of parameters in the flavor breaking sector. 
\subsection{$\epsilon^{\prime}/\epsilon$}
The measure of direct CP violation in the Kaon system is given by the
parameter $\epsilon^{\prime}/\epsilon$. 
Direct CP violation originates from direct transitions of the CP-odd
state into the CP-even $\pi\pi$ final state. The direct CP violation in the neutral 
$K \rightarrow \pi \pi $ decays can be described through the ratio, 
\begin{equation}
\frac{\epsilon^{\prime}}{\epsilon} = e^{i \Phi}
\frac{ w}{\sqrt{2}\left| \epsilon \right| }
\left[ \frac{{\rm Im} A_{2}}{{\rm Re} A_{2}} - 
\frac{{\rm Im} A_{0}}{{\rm Re} A_{0}}  \right]  
 \label{eKprime}
\end{equation}
where $A_{0,2}$ are the isospin amplitudes for the $\Delta I=1/2,3/2$ transitions.
${\rm Im} A_{0,2}$ are calculated from the general low energy effective Hamiltonian
for $\Delta S=1$ transitions \cite{Buchalla:1995vs}. $w= {\rm Re} A_{2}/{\rm Re} A_{0}$ and
$\Phi$ is a strong phase shift difference between the two amplitudes.
In 1999 the NA48 experiment \cite{Fanti:1999nm}
 at CERN and the KTeV \cite{Alavi-Harati:1999xp} 
experiment at FNAL demonstrated that this
observable is actually different from zero as expected in the SM.
The present world average is \cite{epKWA}, 
\begin{equation}
{\rm Re} \left[\frac{\epsilon^{\prime}}{ \epsilon}\right]_{\rm exp} = (16.6 \pm 1.6)\times 10^{-4}. 
 \label{eprimexp}
\end{equation}
There is no simple approximate 
expression for the SM contribution to $\epsilon^{\prime}/\epsilon$.
This calculation is affected by large hadronic uncertainities.
It has been recently pointed out ~\cite{Pich:2004ee}
that to lowest order (in $1/N_{c}$ and in the chiral expansion)
${\rm Re}[ \epsilon^{\prime}/ \epsilon] $ is governed
by the competition between two different decay topologies
and suffers from a strong cancellation between them.
Nevertheless to higher orders chiral loops generate an enhancement of 
the isoscalar amplitude and a reduction of $A_{2}$.
Taking this into account, and following Ref.~\cite{Pich:2004ee}, 
the latest SM prediction is:
\begin{equation}
{\rm Re} \left[ \frac{\epsilon^{\prime}}{\epsilon} \right]_{\rm SM}
= ( 19 
\begin{array}{c}
+ 17 \\
- 18 
\end{array}
) \times 10^{-4}.
 \label{eKprimeSM}
\end{equation}
Therefore even though the SM prediction is 
consistent with the measurement, it does not allow 
us at present to perform stringent 
tests of the CKM mechanism of CP violation.
For a 2003 review of several
calculations see the Refs.~\cite{epsilonprime}.

Since the 1999 measurements, the parameter 
$\epsilon^{\prime}/\epsilon$ has received considerable 
attention in the context of supersymmetric theories \cite{epKsusy}.
The dominant supersymmetric contributions to $\epsilon^{\prime}/\epsilon$
come from the chromomagnetic operators like $O_{g}$,
$O_{g}= g_{s}/(16\pi^{2}) d_{L} \sigma^{\mu \nu}  t^{A}s_{R} G^{A}_{\mu\nu}$.
The Wilson coefficient $C_{g}$  corresponding to this operator is given by,
\begin{equation}
C_{g} = 
(-) \frac{\alpha_{s} \pi}{2 m_{\widetilde{d}}} 
\left[
\frac{4 m_{s}}{m_{\widetilde{d}}}
N(x) (\delta_{\rm LL}^{d})_{12} +
M(x) (\delta^{d}_{\rm LR})_{12}
\right],
\label{Cgepsprime}
\end{equation}
where $N(x)$ and $M(x)$ are dimensionless form
factors defined in Eqs.~\ref{Nxform} and \ref{Mxform} of the appendix.
Taking into account that the relevant hadronic matrix element is given by,
\begin{equation}
\left< (\pi \pi)_{I=0} \left| O_{g} \right| K^{0} \right>  = 
\sqrt{\frac{3}{2}} \frac{11}{16 \pi^{2}} \frac{\left< q \bar{q} \right> }{F_{\pi}^{3}}  m^{2}_{\pi} B_{G},
\label{Og}
\end{equation}
with $F_{\pi} = 131$~MeV and 
where the $B_{G}$ factor is not well known, 
$B_{G}=1-4$ we obtain that the total LR supersymmetric 
contribution to ${\rm Re}[\epsilon^{\prime}/\epsilon]$,
can be conveniently written as,
\begin{equation}
{\rm Re} \left[ \frac{\epsilon^{\prime}}{\epsilon} \right]_{\rm LR+RL}
= \left( 
\frac{ \eta_{\epsilon^{\prime}}}{m_{\tilde{b}}} \right)
\left| {\cal I\rm m} [ (\delta^{d}_{LR})_{21} - (\delta^{d}_{LR})^{*}_{12} ]\right|
N(x),
\label{epspSUSYeta}
\end{equation}
where
\begin{equation}
\eta_{\epsilon^{\prime}} = 
\frac{11 \sqrt{3}}{64 \pi}
\frac{w}{ \left| \epsilon \right| {\rm Re} A^{0}}  
\frac{ m_{K}^{2} m^{2}_{\pi} }{F_{\pi} (m_{s}+m_{d})} 
\alpha_{s}(m_{\widetilde{b}}) \eta B_{G}
\label{etaepsp}
\end{equation}
We used for $w$ and ${\rm Re} A_{0}$
experimental values $w \approx 1/22$ and
${\rm Re} A_{0} = 3.326 \times 10^{-4} $~MeV.
For the rest of parameters we used 
$m_{K}= 490 $~MeV, $m_{\pi}= 140 $~MeV.
$\eta$ is a well known dimensionless strong coupling renormalization factor
defined in Ref.~\cite{Buras:1999da}.
Taking into account the important uncertainities in the current determination
of the $B_{G}$ factor and the lighter quark masses we obtain
the estimate $\eta_{\epsilon^{\prime}} = 100-600$~TeV.
We note that for the model under consideration 
 $$
{\rm Im}[ (\delta_{\rm LR}^{d})_{12}]= {\rm Im}[(\delta_{\rm LR}^{d})_{21}]= 
(-)2\omega s_{\gamma} c_{\beta} \lambda^{3}
 \frac{ v}{m_{\widetilde{b}}} \frac{ \widetilde{m}}{m_{\widetilde{b}}},
  $$
For the worst case scenario, $\eta_{\epsilon^{\prime}}= 600$~TeV, and
assuming that the supersymmetric contribution saturates the experimental measurement
we obtain the constraint,
\begin{equation}
 s_{\gamma} c_{\beta} 
 \left( \frac{ \widetilde{m}}{m_{\widetilde{b}}} \right) 
\left( \frac{ 220~{\rm GeV} }{ m_{\widetilde{b}}} \right)^{2} N(x)
\lesssim {\rm Re} \left[ \frac{\epsilon^{\prime}}{\epsilon} \right]_{\rm exp}.
\label{epspcons}
\end{equation}
For the large $\tan\beta$ case, $\tan\beta=50$ with
$m_{\widetilde{g}} \approx m_{\widetilde{d}}$, 
examined in Sec.~\ref{Radmasses}, the hierarchy
$\widetilde{m}\approx 3m_{\widetilde{b}}$ was required.
In this case using for $\lambda$ and $\omega$ the values determined from the CKM elements
and for $\gamma$, $\gamma=60^{\circ}$ we obtain,
\begin{equation}
m_{\widetilde{b}} \gtrsim 1.3 ~{\rm TeV}~~({\rm \epsilon^{\prime}/\epsilon}).
\label{epspcons2}
\end{equation}
Let us analyze separately the size of the LL and RR contributions.
We see that because of an extra $m_{s}/m_{\widetilde{b}}$
suppression factor the LL+RR contribution to $\epsilon^{\prime}$
is much smaller than the LR contribution.
If the gluino mass is of the same order than the squark masses
(which is required to maximize the loop generated quark masses),
and $\rho \simeq \rho^{\prime}$ 
we obtain a simple approximate expression 
for the LL and RR couplings,  
$$
{\rm Im}[ (\delta_{\rm LL}^{d})_{12}]= {\rm Im}[(\delta_{\rm RR}^{d})_{21}]= 
( 2s_{\gamma} - s_{2\phi^{\prime}} )   \lambda^{3}  \frac{ \widetilde{m}^{2}}{m^{2}_{\widetilde{d}}}
$$
The total LL+RR supersymmetric contribution to ${\rm Re}[\epsilon^{\prime}/\epsilon]$,
can be conveniently written as,
\begin{equation}
{\rm Re} \left[ \frac{\epsilon^{\prime}}{\epsilon} \right]_{\rm LL+RR}
= \left( 
\frac{ 4 m_{s} \eta_{\epsilon^{\prime}}}{m^{2}_{\widetilde{b}}} \right)
{\rm Im} [ (\delta^{d}_{RR})_{21} - (\delta^{d}_{LL})^{*}_{12} ]
M(x).
\label{epsLLRR}
\end{equation}
Using for $\lambda$ the value determined from the CKM elements and
assuming that the supersymmetric contribution saturates the experimental measurement,
we obtain the constraint,
\begin{equation}
 ( 2s_{\gamma} - s_{2\phi^{\prime}} ) 
 \left( \frac{ \widetilde{m}}{m_{\widetilde{b}}} \right)^{2} 
 \left( \frac{ 1.7~{\rm GeV} }{ m_{\widetilde{b}}} \right)^{2}
\lesssim {\rm Re} \left[ \frac{\epsilon^{\prime}}{\epsilon} \right]_{\rm exp}.
\label{epspcons}
\end{equation}
For the large $\tan\beta$ case, $\tan\beta=50$, with
$m_{\widetilde{g}} \approx m_{\widetilde{b}}$ and 
$\widetilde{m}\approx 3m_{\widetilde{b}}$
we obtain $m_{\widetilde{b}} \gtrsim 220$~GeV.
\section{Contributions to CP asymmetries in the B system \label{Bsystem}}
The CP violation measured in neutral K meson decays,
taking into account current experimental and theoretical uncertainities,
can be simply explained with the CKM phase. 
B factories have verified,
especially through measurements of the CP asymmetry 
in the $B_{d} \rightarrow \psi K_{S}$ decay
\cite{BpsiKsBabar,BpsiKsBelle},
that the CP symmetry is significantly violated in the B sector, in agreement with Standard
Model predictions, providing a confirmation of the so called CKM paradigm \cite{CKM,Ligeti:2004ak}. 
This fact does not
rule out the possibility that the effects of CP phases of a different origin,
as for instance the phases in the soft supersymmetry breaking sector
\cite{Ciuchini:1997zp,Barbieri:1997kq,Masiero:2001cc},
could manifest in the near future
\cite{Browder:2004wu} through other CP violating observables, especially penguin 
dominated amplitudes such as $B\rightarrow \phi K^{0},\eta^{\prime} K_{s}$. 
In this section we will study the constraints that the currents
measurements of CP asymmetries in several $B$-decays imposse
on the model under consideration. 
\subsection{CP asymmetry in $B \rightarrow \psi K_{S}$ \label{BetaKs}}
The $B^{0}\bar{B^{0}}$ mixing amplitude is defined by
the matrix element of the effective $\Delta B=2$ hamiltonian
as ${\cal M}_{b}= M_{12}(B^{0}_{d}) = 
\left< \bar{B}_{d} \left| {\cal H}_{\rm eff}^{\Delta B=2} \right| B_{d} \right> $.
The phase of the mixing amplitude is related with the mixing CP asymmetry in 
the decay $B\rightarrow \psi K_{s}$ by,
\begin{equation}
S_{\psi K_{s}} = \sin \left[ {\rm Arg} \left[ {\cal M}_{b} \right] \right].
\end{equation}
According to the most recent averaged experimental
results of Babar and Belle 
$S_{\psi K_{s}} = 0.736 \pm 0.049$.
This can be simply accounted to date with the CKM phase. 
If future measurements reduce considerably the experimental 
uncertainity in $S_{\psi K_{S}}$, there is hope that deviations from the SM prediction could
be elucidated. 
It is convenient to 
separate the SM and supersymmetric contributions to the
mixing amplitude in the form ${\cal M}_{b} = 
{\cal M}_{b}^{\rm SM}+ {\cal M}_{b}^{\rm SUSY}$.
We also find convenient to define 
$\phi_{1}$ and $\phi_{1}^{\prime}$ as the
phases of the SM and supersymmetric component 
of the amplitude respectively, {\it i.e.}
${\cal M}_{b}^{\rm SM} = e^{i2\phi_{1}} |{\cal M}_{b}^{\rm SM}|$ and 
${\cal M}_{b}^{\rm SUSY} = e^{i2\phi_{1}^{\prime}} |{\cal M}_{b}^{\rm SUSY}|$.
Since the SM prediction can account perfectly for the experimental result
we expect that the supersymmetric contribution is a small correction and
expand the expression for the CP asymmetry 
in powers of the ratio $R_{\psi} =  |{\cal M}_{b}^{\rm SUSY}| 
/ |{\cal M}_{b}^{\rm SM}|$. To leading order,
\begin{equation}
S_{\psi K_{s}} = 
\sin 2\phi_{1}  [ 1  - \sin (2(\phi_{1} - \phi_{1}^{\prime}))  R_{\psi} ]
+ \sin(2\phi_{1}^{\prime})  R_{\psi} \label{SphiKNP}.
\end{equation}
It is known that in the absence of new physics contributions 
the SM CP phase can account for the
present experimental results for $S_{\psi K_{s}}$.
From Eq.~\ref{SphiKNP} we can obtain constraints on
$R_{\psi}$ and $\phi_{1}^{\prime}$.
Assuming that the new physics contribution saturates 
a 50\% of the experimental uncertainity
we obtain to leading order in $ R_{\psi} $,
\begin{equation}
R_{\psi} 
\sin(2(\phi_{1}^{\prime}-\phi_{1})) \lesssim 0.5
\label{psiKcons2}
\end{equation}
We note that even in the limit where the complex phase of the SUSY amplitude goes to
zero the mixing CP asymmetry, $S_{\psi K_{S}}$,
is affected by the SUSY contributions through their effects on the absolute 
value of the amplitude, $ S_{\psi K_{s}} = 
\sin 2\phi_{1}^{\rm SM}  [ 1  - \sin (2\phi_{1}^{\rm SM})  R_{\psi} ]$.
The mass difference in the $B_{d}\bar{B}_{d}$ system, $\Delta m_{d}
= 2 {\rm Abs}[ {\cal M}_{b}]$, 
also puts a stringent constraint on $R_{\psi}$.
$\Delta m_{d}$  is an observable well known experimentally,
to the level of $1.5$\%.
The experimental measurement yields 
$\Delta m_{d} = (3.22 \pm 0.05 ) \times 10 ^{-10}$~MeV.
The SM prediction for $\Delta m_{d}$ is about
$\Delta m_{d}^{\rm SM} = (2.9\pm 2.2)\times 10 ^{-10}$~MeV.
We note that even tough the theoretical uncertainity is about $75$\%
the central value is only $10$\% from the central experimental value.
Assuming that the supersymmetric contribution saturates 
a $50\%$ of the experimental measurement and
expanding ${\cal M}_{b}$ in powers of $R_{\psi}$ 
we obtain the constraint,
\begin{equation}
 R_{\psi} 
   \cos(2(\phi_{1}^{\prime}-\phi_{1})) \lesssim 0.5,
\label{dmdcons2}
\end{equation}
We note that we have assumed that 
$(\phi_{1} - \phi_{1}^{\prime}) \neq \pm \pi /4$.
If this was not the case, then the 
second order term in the expansion would be dominant and
we would obtain a milder constraint,
$R_{\psi} ^{2}
\lesssim  0.5$. 

Next we need to calculate the expressions for $R_{\psi}$ and $\phi_{1}^{\prime}$ in our model.
The previous constraints in Eq.~\ref{psiKcons2} and \ref{dmdcons2} will
translate into lower bounds on the squark mass spectra. 
Again, we identify four qualitatively different contributions to  
$ {\cal M}_{b}^{\rm SUSY}$,
that in our model, working to leading order in powers of $\lambda$, 
take the following values, 
\begin{eqnarray}
 (\delta_{LR}^{d})^{2}_{13}+(\delta_{RL}^{d})^{2}_{13}  &=& 
2 \delta_{\rm LR}^{13} (3+2c^{2}_{\gamma} - 2 i s_{2\gamma}) 
\\
(\delta_{LR}^{d})_{12} (\delta_{RL}^{d})_{12} &=& \delta_{\rm LR}^{13}
( 2 c_{\gamma}^{2} - 5 - i s_{2\gamma})
\label{deltasLR13}
 \end{eqnarray}
where,
\begin{equation}
\delta_{\rm LR}^{13} =   \lambda^{4}  c^{2}_{\beta}
 \frac{ v^{2}}{m^{2}_{\widetilde{b}}}  \frac{\widetilde{m}^{2}}{m^{2}_{\widetilde{b}}}, 
\end{equation}
and,
\begin{eqnarray}
  (\delta_{LL}^{d})^{2}_{13}+(\delta_{RR}^{d})^{2}_{13} &=& 
 2 \delta_{\rm LL }^{13} e^{2i(\phi-\gamma)} ( 1 + 4e^{2i\gamma}),
  \\
(\delta_{LL}^{d})_{13}(\delta_{RR}^{d})_{13} &=& \delta_{\rm LL}^{13}
  e^{2i(\phi-\gamma)} ( 1 - 4e^{2i\gamma}),
\label{deltasLL13}
\end{eqnarray}
where,
\begin{equation}
\delta_{\rm LL}^{13} = \rho^{\prime 2} \lambda^{4}.  
\label{deltaLLRR13}
\end{equation}
We see that in this model the LL and RR delta couplings
in general are expected to dominate 
the contribution to ${\rm Im}[{\cal M}_{b}^{\rm SUSY}]$
since the LR couplings contain an additional suppression factor,
$c^{2}_{\beta}  v^{2}/m^{2}_{\widetilde{b}}$.
It is possible to add all the dominant 
$\delta_{LL}$ and $\delta_{RR}$ contributions
to the Wilson coefficients to give a simple approximate 
expression
for the supersymmetric contribution to ${\cal M}_{b}$.
We obtain, 
\begin{eqnarray}
{\cal M}_{b}^{\rm SUSY} &=&  
\frac{\alpha^{2}_{s}m_{B} f^{2}_{B}  X^{2}_{B}}{60 m^{2}_{\widetilde{b}}} \left[
 \left( (\delta_{LL}^{d})^{2}_{13} + (\delta_{RR}^{d})^{2}_{13} \right) \right. 
 \nonumber \\
&& \left. \times D(x) + (\delta_{LL}^{d})_{13} (\delta_{RR}^{d})_{13} C(x) \right]
\label{M13LLRR}
 \end{eqnarray}
Here $X^{2}_{B}$ is a dimensionless factor defined 
as $X^{2}_{B}=m^{2}_{B}/(m_{b}(m_{b})+m_{d}(m_{b}))^{2}$,
numerically $X_{B}\approx 1.08$.
$x$ is defined as $x= m^{2}_{\tilde{g}}/m^{2}_{\widetilde{b}}$.
$C(x)$ and $D(x)$ are dimensionless form factors which were already
introduced previously in Sec.\ref{epsilon}
as $C(x)=C_{f}f(x)+C_{g}g(x)$ and $D(x)= D h(x)$.
 $f(x)$, $g(x)$ and $h(x)$ are dimensionless form factors defined in 
Eqs.~\ref{fx}, \ref{gx} and \ref{hx} of the appendix.
The constant coefficients $C_{f}$, $C_{g}$ and $D$,
like in the $\Delta S=2$ case, 
absorb the dependency on the method of calculation of the hadronic matrix elements
as well as the renormalization effects on the Wilson 
coefficients. We have evaluated
$C_{f}$, $C_{g}$ and $D$ 
following Ref.~\cite{Becirevic:2001jj} 
where lattice QCD methods were used to calculate the
relevant hadronic matrix elements. 
We obtain that $D\ll C_{g} \ll C_{f}$ and 
$C_{f } \approx 7.33$. 
Had we used the naive vacuum insertion approximation at tree level 
we would obtain,
$C_{f}   =  (48 X_{B}^{2}  +9)/(27 X_{B}^{2})  \approx  2.06$
which would imply an 
underestimation of the dominant term by a factor of order $1/10$.
The coefficients $C_{f}$,$C_{g}$ and $D$ 
also depend on the scale of the supersymmetric spectra. We
have calculated our numerical values at the scale $M_{S}= 1$~TeV
using the renormalization factors given in Ref.~\cite{Becirevic:2001jj}.
The contribution of the form $\delta_{LL} \delta_{RR}$  
in Eq.~\ref{deltaLLRR13} clearly dominates the
supersymmetric contribution. We use
the well known expression for the SM contribution to
the amplitude ${\cal M}_{b}$, see for instance Eq.(3.60) in Ref.~\cite{Buras:2001pn},
\begin{equation}
{\cal M}_{b}^{\rm SM} = 
\frac{ G_{F}^{2}}{12 \pi^{2}} \eta_{B} \hat{B}_{B_{d}} f_{B}^{2}m_{B_{d}}m_{W}^{2} 
\left|V_{td}V_{tb}^{*}\right|^{2} S(x_{t}),
\end{equation}
where $\eta_{B} = 0.55(1)$ is a QCD correction factor,
$\hat{B}_{B_{d}}$ is a renormalization group invariant
parameter available in the literature
\cite{Sachrajda:2000ci}, $\hat{B}_{B_{d}}=1.30\pm0.12$, and
$S(x_{t})$ is a dimensionless form factor given by
$S(x_{t}) = 2.46 (m_{t}/170~{\rm GeV})^{1.52}$.
We can write the ratio of the supersymmetric contribution 
over the SM amplitude in the form,
\begin{equation}
\left.  R_{\psi} \right|_{\rm LL+ RR} 
\approx   
 \frac{\eta^{2}_{\psi}}{ m^{2}_{\widetilde{b}}} \rho^{\prime2}\lambda^{4} f(x)
 (17 - 8c_{2\gamma})^{1/2},
 \label{Rpsi}
 \end{equation}
where,
\begin{equation}
\eta_{\psi} \approx
 \frac{\alpha_{s} \pi X_{B} C_{f}^{1/2}}{
\sqrt{5} G_{F } \left|V_{td} V_{tb}\right| B_{B}^{1/2} \eta_{B}^{1/2} m_{W} S^{1/2}(x_{t})} \approx 78~\hbox{TeV}.
\end{equation}
We will substitute in the expression \ref{Rpsi},
the value of $\lambda$ determined from the quark data and
for $\gamma$ we will use the central value of the $1\sigma$ global fit. 
$\gamma=60^{\circ}$. For the large $\tan\beta$ case, $\tan\beta=50$,
examined in Sec.~\ref{Radmasses},
assuming that $\rho^{\prime}\simeq \rho \approx 9$
and $m_{\widetilde{g}}\approx m_{\widetilde{b}}$ 
the constraints in Eq.~\ref{psiKcons2}
and \ref{dmdcons2} reduce to,
\begin{eqnarray}
\left( \frac{700~\rm TeV}{ m_{\widetilde{b}}} \right)^{2}
 c_{2(\phi_{1}^{\prime}-\phi_{1})}  &\lesssim& 0.5, 
 \label{strongest} 
 \\
\left( \frac{700~\rm TeV}{ m_{\widetilde{b}}} \right)^{2}
 s_{2(\phi_{1}^{\prime}-\phi_{1})}  &\lesssim& 0.13,
\end{eqnarray}
The phase $\phi_{1}^{\prime}$ can be calculated from Eq.~\ref{deltasLL13}.
We obtain, 
\begin{equation}
\tan(2\phi_{1}^{\prime}) = \frac{ 4 s_{2\phi} + s_{2(\gamma-\phi)}}{4c_{2\phi}- c_{2(\gamma-\phi)}}.
\end{equation}
The phase $\phi$ is not constrained by the quark masses and mixings. 
The value of $\phi_{1}^{\prime}$ depends strongly on the phase $\phi$.
For $\phi=0$ and $\gamma=60^{\circ}$ we obtain $\phi_{1}^{\prime}=10.9^{\circ}$
while for $\phi=30^{\circ}$ and $\gamma=60^{\circ}$ we obtain $\phi_{1}^{\prime}=70.9^{\circ}$.
When $\cos(2(\phi_{1}^{\prime}-\phi_{1}))\approx 1$
the strongest phase independent constraint on the squark spectra comes from Eq.~\ref{strongest}, 
\begin{eqnarray}
m_{\widetilde{b}} &
\gtrsim& 1000~ {\rm TeV}~~~(\Delta m_{d} \quad \phi_{1}^{\prime}\approx \phi_{1}), \\
m_{\widetilde{b}} &\gtrsim& 1000~ {\rm TeV}~~~(S_{\psi K_{s}} \quad \phi_{1}^{\prime}\approx \phi_{1}\pm \pi/2). 
\end{eqnarray}
On the other hand if $\rho^{\prime} \ll 1$ 
these constraints on the squark spectra would be milder.
This will happen for instance when $\eta^{\prime} \ll \eta$.
\subsection{CP asymmetry in $B \rightarrow \phi K_{S}$ \label{BphiKs}}
The latest results from BELLE collaboration 
for the time dependent CP asymmetry coefficient $S_{\phi K}$ derived from
the combined $\phi K^{0}$ dataset are \cite{BphiKsBelle}
$S_{\phi K}^{\rm BELLE} = + 0.06 \pm 0.42$ while
the latest results from BABAR collaboration for the same coefficient are \cite{BphiKsBabar}
$S_{\phi K}^{\rm BABAR} = + 0.50 \pm 0.32 $.
Combining the results from both experiments one obtains the world average,
$S_{\phi K}^{\rm BABAR+BELLE} = + 0.34 \pm 0.21$ \cite{Ligeti:2004ak}.
Taking into account that 
the SM prediction for the time dependent CP asymmetry is
$S_{\phi K}= \sin (2\phi_{1})=0.726\pm0.037$ (here we used the world averaged
CP asymmetry determined from charmonium final states) 
the current world average
seems to differ from the SM expectation by about $2\sigma$ level.
Therefore this process is one of the best candidates 
for the manifestation of new physics in the quark sector.
$S_{\phi K}$ in the context of supersymmetric theories has
received considerable attention recently \cite{Ciuchini:1997zp,Barbieri:1997kq,susyBphiK02,Khalil:2002fm,Ciuchini:2002uv,Kane:2004ku,Baek:2003kb,Chakraverty:2003uv,Khalil:2003ng,Wang:2003dw,Mishima:2003ta,Dutta:2003hb,susyBphiK03,SUSYphiK04,Gabrielli:2004yi,Khalil:2004wp}
since any clear deviation from the SM prediction
would imply the existence of new CP phases other than the CKM phase.

The total decay amplitude, $A_{\phi K_{S}} = 
< \phi K_{S} | 
{\cal H}_{\Delta B=1}^{\rm eff \dagger} | B^{0} >$
can be written in the form, $A_{\phi K_{S}} = 
A_{\phi K_{S}}^{\rm SM}  + A^{\rm SUSY}_{\phi K_{S}}$.
Additionally one can parametrise the SM and supersymmetric contributions
to the amplitude in the form,
$A^{\rm SUSY}_{\phi K_{S}} = | A^{\rm SUSY}_{\phi K_{S}} |
 e^{i \phi_{\rm NP}} e^{i \delta_{\rm SUSY}}$ and
$A^{\rm SM}_{\phi K_{S}}= 
| A^{\rm SM}_{\phi K_{S}} | e^{i \delta_{\rm SM}}$.
where $\phi_{\rm NP}$ is the CKM-like complex phase of the supersymmetric contribution
and $\delta_{\rm SM}$ and $\delta_{\rm SUSY}$
are the SM and supersymmetric CP conserving strong phases
respectively.
Assuming that the susy contribution to the amplitude is smaller than the SM one,
and expanding in powers of the ratio $R_{\phi} =
| A_{\phi K_{S}}^{\rm SUSY} |/| A^{\rm SM}_{\phi K_{S}} |$ 
it is possible to obtain
approximate expressions for the direct and mixing CP asymmetries
\cite{Gronau:1992rm,Khalil:2002fm}. To leading order in $R_{\phi}$, 
\begin{equation}
S_{\phi K_{S}}  =  s_{2\phi_{1}} + 2 s_{\phi_{\rm NP}}c_{\delta} c_{2\phi_{1}} 
R_{\phi}  ,
\label{EqSphi}
\end{equation}
where $\delta$ is the difference of supersymmetric and SM CP conserving strong phases,
$\delta = (\delta_{\rm SM} - \delta_{\rm SUSY})$. 
We will constraint the supersymmetric contribution to $S_{\phi K_{S}}$ 
assuming that this contribution accounts for the difference between
the experimental measurement and the SM prediction, i.e.,
\begin{equation}
2 s_{\phi_{\rm NP}} c_{\delta} c_{2\phi_{1}} R_{\phi} 
\lesssim 
 , 0.40 \pm 0.26
\label{consSphi}
\end{equation}
where for $s_{2\phi_{1}}$ we used the experimental 
value of $S_{\psi K_{S}}$, $S_{\psi K_{S}}=0.736\pm 0.049$.
There are two basic contributions to the supersymmetric amplitude: the contributions
coming from the $\delta_{\rm LR}$ couplings and the contributions coming from 
the $\delta_{LL}$ and $\delta_{RR}$ couplings.
We will first obtain a simple expression for the $\delta_{\rm LR}$ contributions.
We note that at the susy scale there is only
one Wilson coefficient which contains the coupling $(\delta_{LR}^{d})_{23}$,
the chromomagnetic operators like $O_{g}$,
$O_{g}= g_{s}/(16\pi^{2}) d_{L} \sigma^{\mu \nu}  t^{A}b_{R} G^{A}_{\mu\nu}$.
The Wilson coefficient $C_{g}$  corresponding to this operator at the susy scale 
is given by,
\begin{equation}
C_{g}(m_{\widetilde{b}}) = 
(-) \frac{\alpha_{s} \pi}{2 m_{\widetilde{b}}} 
\left[
\frac{4 m_{b}}{m_{\widetilde{b}}}
N(x) (\delta_{\rm LL}^{d})_{23} +
M(x) (\delta^{d}_{\rm LR})_{23}
\right],
\label{CgBphiKs}
\end{equation}
Here $x$ is defined as $x= m^{2}_{\tilde{g}}/m^{2}_{\tilde{d}}$.
$M(x)$ and $N(x)$ are invariant dimensionless form factors that we have 
conveniently normalized so that  $M(x),N(x) \approx 1$ when $x\rightarrow 1$,
see Eqs.~\ref{Nxform} and \ref{Mxform} of the appendix.
When calculating the supersymmetric contributions to the asymmetry
one has to take into account the renormalization of the Wilson coefficients
from the susy scale down to the bottom mass scale.
Following Ref.~\cite{Ali:1997nh}
we have included the NLO corrections using the 
generalized factorisation approach assuming that
$m_{\widetilde{b}}\simeq 1$~TeV. We note that all the 
$\delta_{LR}$ contributions to the low energy effective Wilson coefficients
arise originally from the Wilson coefficient $C_{g}(m_{\widetilde{b}})$ in Eq.~\ref{CgBphiKs}.
Therefore all the contributions to each effective Wilson coefficient coming from 
the flavor violating soft susy breaking trilinear couplings ({\it i.e.} $\delta_{LR}$ couplings) 
can be added up since they are proportional to the same gluino-squark form factor $M(x)$.
The resulting contribution can always be written in the form, 
\begin{equation}
\left. A^{\rm SUSY}_{\phi  K_{S}} \right|_{\rm LR+RL}
 = \frac{ \alpha_{s}^{2} f_{\phi K}  M(x)}{18 m_{b} m_{\widetilde{b}} }
\left( (\delta^{d}_{\rm LR})_{23} + (\delta^{d}_{\rm LR})^{*}_{32} \right) .
\label{AphiKLR}
\end{equation}
In our notation the coefficient $f_{\phi K}$ absorbs the dependency on the
method of calculation of the hadronic matrix elements. 
We have calculated $f_{\phi K} $ using
the generalized factorisation approach, following Ref.~\cite{Ali:1997nh}.
$f_{\phi K} $ is parametrized in the form $f_{\phi K} = f_{q} X_{\phi}$, 
$f_{q}$ is a factor associated with the momentum carried by the gluon  
in the corresponding penguin diagram, $f_{q}= m_{b}/\sqrt{\left<q^{2}\right>}$
( in the rest of the paper we will assume that $f_{q}=\sqrt{2}$). 
$X_{\phi}$ arises from the hadronic matrix elements. It is given by
$X_{\phi} = 2 F_{1}^{B\rightarrow  K}( m^{2}_{\phi}) 
f_{\phi} m_{\phi} (p_{K}\cdot \epsilon_{\phi} )$. The numerical value
of the parameter $X_{\phi}$ is irrelevant for our purposes because
it cancels with the same factor coming from the SM contribution.
The coefficient $f_{\phi K}$ could be calculated using 
other more recent and precise approaches 
which are available in the literature: as the perturbative QCD approach \cite{pQCD} or 
the QCD factorisation approach \cite{BBNS}. Nevertheless
it has been pointed out that in that case one 
would obtain sligthly different values for the relevant coefficients
\cite{Mishima:2003ta,Dutta:2003hb,Kane:2004ku,Gabrielli:2004yi}.
Therefore for our purpose, which is to obtain a good
estimate of the constraint on the new physics contributions,
the generalized factorisation approach is precise enough.

The contributions to the supersymmetric amplitude
coming from the $\delta_{\rm LL}$ and $\delta_{\rm RR}$ couplings
can also be written in a similar fashion, 
\begin{equation}
\left. A^{\rm SUSY}_{\phi K_{S}}\right|_{\rm LL+ RR}
 =  \frac{  \alpha_{s}^{2} X_{\phi} L(x) }{47 m^{2}_{\tilde{g}}} 
\left( (\delta^{d}_{\rm LL})_{23}  +  (\delta^{d}_{\rm RR})_{23} \right)  .
\label{AphiKLL}
\end{equation}
$L(x)$ is a dimensionless polynomial conveniently normalized such that
$L(x) \rightarrow 0$ when $x\rightarrow 0$ and
$L(x) \approx 1$ when $x\rightarrow 1$. 
We note that in this case, since there are $\delta_{\rm LL}$ and $\delta_{RR}$ 
contributions coming from different Wilson coefficients,
the coefficients of the polynomial $L(x)$ 
depend on the method used to evaluate the hadronic matrix elements
and to a lesser extent on the scale of the supersymmetric spectra but they 
do not depend on the flavor mixing structure in the susy breaking sector.
We have evaluated the coefficients of $L(x)$ numerically 
using the generalised factorisation approach following Ref.~\cite{Ali:1997nh}.
We obtain approximately,
\begin{equation}
L(x)_{\rm GF} = c_{0} + c_{1} (x-1)  ,\quad 
c_{0} \approx 1, ~c_{1} \approx -\sqrt{3},
\end{equation}
in the limit $x\rightarrow 1$.
Additionally, if one is interested in the limit 
$x\simeq 0$, {\it i.e.} $m_{\tilde{g}}\ll m_{\tilde{q}}$,
$L(x)$ is approximated by,
\begin{equation}
L(x)_{\rm GF} = (-) 4 x \left( d_{0} + d_{1} \ln x \right), \quad 
d_{0} \approx 18, ~ d_{1} \approx 7 ,
\end{equation}
where the coefficients $c_{0,1}$ and $d_{0,1}$ shown here 
are good approximations to the actual values calculated numerically. 
The expressions in Eqs.~\ref{AphiKLR} and \ref{AphiKLL} are practical expressions
of general interest irrespective of the form of the matrices
$\delta_{LR}$, $\delta_{\rm LL}$ and $\delta_{\rm RR}$.
We see from Eqs.~\ref{AphiKLR} and \ref{AphiKLL}
that one naively expects that the $\delta_{\rm LR}$ contributions 
dominate since the $\delta_{LL}$ and $\delta_{RR}$ contributions 
receive in general an additional suppression factor 
of the order $m_{b}/(5m_{\widetilde{b}})$. Nevertheless
for the model under consideration we obtain,
\begin{eqnarray}
(\delta^{d}_{\rm LR})_{23} + (\delta^{d}_{\rm LR})^{*}_{32}  &=& 
4 \lambda ( c_{\phi} \omega - \frac{\widetilde{m}}{A_{b}}) \frac{v A_{b}}{m^{2}_{\widetilde{b}}}c_{\beta},
\\
(\delta^{d}_{\rm LL})_{23}  +  (\delta^{d}_{\rm RR})_{23}   &=& 
-4 e^{i\phi} \rho^{\prime} \lambda.
\label{dLLRR23}
\end{eqnarray}
We note that the total $\delta_{LR}$ contribution to $S_{\phi K_{S}}$ is zero since
${\cal I}{\rm m}[ (\delta^{d}_{\rm LR})^{*}_{23} + (\delta^{d}_{\rm LR})_{32} ]=0$.
Therefore in our model  
we find that only (LL+RR) couplings contribute to $S_{\phi K_{S}}$. 
We find the following simple expression for the ratio of the dominant
supersymmetric contribution to the amplitude 
over the Standard Model contribution,
\begin{equation}
\left. R_{\phi} \right|_{\rm LL+RR}=
\left(
\frac{ \eta^{2}_{\phi}}{m^{2}_{\tilde{q}}} \right)
L(x)  \left| (\delta^{d}_{\rm LL})_{23}  +  (\delta^{d}_{\rm RR})_{23} \right|
\label{BphiKsdom}
\end{equation}
where $\eta_{\phi}$ 
is a coefficient independent of the supersymmetric parameter space given by, 
\begin{equation}
\eta_{\phi}^{2} \approx
\frac{ \sqrt{2}  \alpha_{s}^{2} }{45 G_{F} 
\left| V_{tb}^{*} V_{ts} \right| h_{\phi}} \approx (189~{\rm GeV})^{2}. 
\label{etaphi}
\end{equation}
Here $h_{\phi}$ 
parametrizes the dependence of the 
SM contribution on the Wilson coefficients and
hadronic matrix elements. We used the value for $h_{\phi}$ calculated numerically
in Ref.~\cite{Ali:1997nh} using the generalised factorization approach (GF) .
For instance if
$m_{\tilde{q}}= 500$~GeV we obtain $\left| R_{\phi} \right|_{\rm LL+RR}  
\approx 0.14 L(x) \left| (\delta_{LL}^{d})_{23} + (\delta_{RR}^{d})_{23} \right|$,
which agrees with previous numerical calculations \cite{Khalil:2002fm,Khalil:2004wp}.
We note that 
Eq.~\ref{BphiKsdom} provides some analytical insight in the dependency of the
supersymmetric contributions on the supersymmetric spectra,
especially on the gluino squark mass ratio through the form factor $L(x)$.

Finally we will use the expression for $((\delta_{LL}^{d})_{23} + (\delta_{RR}^{d})_{23} )$
in our model given in Eq.~\ref{dLLRR23} and the general 
expression for the amplitude $A_{\phi K_{S}}$ given in 
Eq.~\ref{BphiKsdom} to rewrite the constraint from Eq.~\ref{consSphi}.
Using for $\lambda$, $\lambda\approx 0.22$, we obtain,
\begin{equation}
\rho^{\prime} s_{\phi_{\rm NP}} c_{\delta} \left( \frac{212~{\rm GeV}}{m_{\widetilde{b}}} \right)^{2} 
L(x) \lesssim 0.40 \pm 0.26
\label{phiKscons}
\end{equation}
We note that the phase $\phi_{\rm NP}$ as well as the strong phases difference $\delta$
are not constrained by the data on quark masses
and mixings. If $\phi_{\rm NP}=0$ this contribution to the asymmetry $S_{\phi K}$ 
would cancel.
Let us assume in the worst case scenario that
$\phi_{\rm NP}=\pi/2$, $\delta=\pi/2$ and $\rho^{\prime}=\rho=9$ 
(which is the value for the large $\tan\beta$ scenario analyzed in Sec.~\ref{Radmasses}).
We would obtain only a mild constraint
on the squark mass scale, of the order $m_{\widetilde{b}} \gtrsim 1~$TeV.

Finally we would like to mention that, 
as it has been pointed out, in the case when the $\delta_{\rm LR}$ contribution is much smaller than 
the $\delta_{LL}$ or $\delta_{RR}$ contributions,
the chargino contributions to the amplitude may be relevant
since they could be  
of the same order than the $\delta_{LL}$ and $\delta_{RR}$ 
gluino contributions 
\cite{Baek:2003kb,Chakraverty:2003uv,Khalil:2003ng,Gabrielli:2004yi,Khalil:2004wp,Wang:2003dw}.
A more precise calculation would require the inclusion of these contributions.
\subsection{CP asymmetry in $B \rightarrow \eta^{\prime} K_{S}$ \label{BetaKs}}
Recent results on the measurements of the CP asymmetries 
on the $b\rightarrow s$ processes have 
reported possible anomalies not only in $B\rightarrow  \phi K_{s}$ 
but also in other processes, including $B \rightarrow \eta^{\prime} K_{s}$.
The latest results from BELLE \cite{BphiKsBelle}
and BABAR \cite{BphiKsBabar} collaborations 
for the time dependent CP asymmetry coefficient $S_{\eta^{\prime} K_{s}}$,
$(S_{\eta^{\prime} K}^{\rm BELLE}) \stackrel {\rm 2004}{=} + 0.06 \pm 0.42$ 
and $ (S_{\eta^{\prime} K}^{\rm BABAR})\stackrel{\rm 2004}{=} + 0.50 \pm 0.32$,
seems to differ from the SM expectation. 
Combining the results from both experiments one obtains the world average,
$(S_{\eta^{\prime} K}^{\rm BABAR+BELLE})\stackrel{\rm 2004}{=} + 0.41 \pm 0.11$ \cite{Ligeti:2004ak}.
This has motivated the recent interest in the supersymmetric contributions 
to the CP asymmetry in the decay  $B \rightarrow \eta^{\prime} K_{s}$
versus $B \rightarrow \phi K_{s}$  \cite{recenteta,Khalil:2004wp}
as well as in correlations with other supersymmetric processes \cite{Gabrielli:2004yi,Datta:2004jm}.
It is known that because vector mesons $(\phi, \rho,\cdot \cdot \cdot )$ and
pseudoescalar mesons $(\pi, K, \eta^{\prime} ,\cdot \cdot \cdot )$
have opposite parity the B decays
to these two final states will be sensitive to different combinations
of the relevant Wilson coefficients \cite{Kagan:1998bh}. For instance,
in supersymmetric theories
the gluino loop effects 
coming from $\delta_{\rm LR}$ couplings will
contribute by a factor proportional to 
$((\delta_{\rm LR})_{23} +(\delta_{\rm LR})^{*}_{32})$
in the vector case and to $((\delta_{\rm LR})_{23} - (\delta_{\rm LR})^{*}_{32})$
in the pseudoscalar case respectively.
For the model under consideration,
the contributions from $\delta_{\rm LR}$ couplings,
which exactly cancel for the $S_{\phi K_{s}}$ asymmetry, 
not only do not cancel but 
dominate the CP asymmetry in the decay $B\rightarrow \eta^{\prime} K_{s}$.
We can obtain an 
expression for the $\delta_{\rm LR}$ contribution to $A_{\eta^{\prime}K_{s}}$
similar to $A_{\phi K_{s}}$ in 
Eq.~\ref{AphiKLR} with the change 
$((\delta_{\rm LR})_{23} +(\delta_{\rm LR})^{*}_{32})
\rightarrow ((\delta_{\rm LR})_{23} - (\delta_{\rm LR})^{*}_{32})$. Using the
resulting expression we obtain the following 
simple formula
for the supersymmetric contribution to $S_{\eta^{\prime} K_{s}}$,
\begin{equation}
S_{\eta^{\prime} K_{s}} \approx 
4 c_{\beta} \omega \lambda s_{\phi}  M(x)
\left( \frac{3.4~{\rm TeV}}{m_{\widetilde{b}}} \right)^{2}
\left( \frac{A_{b}}{m_{\widetilde{b}}} \right) 
\label{etaKsLR}
\end{equation}
Using the values of $\lambda$ and $\omega$ determined
from quark masses and mixings, and assuming that $A_{b}\approx m_{\widetilde{b}}$
we obtain the following constraint on 
$S_{\eta^{\prime} K}$,
\begin{equation}
\left| s_{\phi} c_{\beta} c_{2\phi_{1}} s_{\delta} \right| \left( \frac{1~{\rm TeV}}{m_{\widetilde{b}}} \right)^{2} 
M(x) \lesssim 0.33 \pm 0.16.
\label{phiKscons}
\end{equation}
We note that the phase $\phi$ as well as $\delta$, the difference between 
strong phases, are not constrained by the data.
 If $\phi\approx 0$ this contribution to the asymmetry would cancel.
In the worst case scenario, assuming that $x \simeq 1$,
$\phi=\pi/2$ and $\delta=\pi/2$ the constraint depends strongly 
on the value of $\tan\beta$.  For large $\tan\beta$, $\tan\beta=50$,
we would obtain a mild lower constraint
on the squark mass scale, $m_{\widetilde{b}} \gtrsim 250~$GeV.
\subsection{CP asymmetry in $B\rightarrow X_{s} \gamma$
\label{BXsgamma}}
CLEO collaboration has set a range on the direct CP asymmetry in the 
$b \rightarrow s \gamma$ decay, 
$A_{\rm CP}^{b\rightarrow s \gamma}$,
at $90~$\% C.L. as $A_{\rm CP}^{b\rightarrow s \gamma}=(-3.5 \pm 13.5) \%$
\cite{Aubert:2004hq}
while the BELLE collaboration also set a range as 
$A_{\rm CP}^{b\rightarrow s \gamma}=(-0.8 \pm 10.7) \%$  \cite{Nishida:2003yw}. 
According to the SM theoretical prediction 
$A_{\rm CP}^{b\rightarrow s \gamma}$ is smaller than $1\%$
\cite{Soares:1991te}.
Therefore $A_{\rm CP}^{b\rightarrow s \gamma}$ is 
an observable potentially sensitive to the presence of new physics.
Furthermore it is expected that the experimental uncertainity will be reduced
to less than 1\% at a super B factory.
$A_{\rm CP}^{b\rightarrow s \gamma}$ in supersymmetric theories has
received considerable interest recently \cite{CPbsgaSUSYrec,Gabrielli:2004yi,Datta:2004jm,Ciuchini:2002uv}.
It is known that 
a CP violating phase in the entries $(\delta_{\rm LR, RL}^{d})_{23}$
or  $(\delta_{\rm LL, RR}^{d})_{23}$
will generate CP violation in the decay $B \rightarrow X_{s}\gamma$
\cite{CPbsgammaSUSY,Kagan:1998bh}.  
The direct CP asymmetry in $b\rightarrow s\gamma$ decay can be written
in terms of  the effective Wilson coefficients 
of the low-energy effective weak Hamiltonian \cite{Kagan:1998bh},
\begin{eqnarray}
A_{\rm CP}^{b\rightarrow s \gamma}  &=& 
\frac{1}{\left|C^{L}_{7}\right|^{2}+\left|C^{R}_{7}\right|^{2}}
\left[ a_{27}~ {\rm Im} [ C_{2}( C^{L*}_{7}+ C^{R*}_{7})]\right.
\nonumber
\\
& &
+ a_{g7} ~ {\rm Im} [ C^{L}_{g} C^{L*}_{7} 
 + C^{R}_{g} C^{R*}_{7}]   
 \nonumber\\
&& + \left.
a_{2g} ~{\rm Im} [ C_{2}( C^{L*}_{g}+C^{R*}_{g})] \right].
\label{AXsgamma}
\end{eqnarray}
where $C^{L}_{7} = C^{L\rm eff}_{7}(m_{b})$, $C^{L}_{g} = C_{g}^{L \rm eff}(m_{b})$
and $C_{2}$ multiply the chromo-magnetic dipole operators, 
$O_{7} = \frac{e}{16 \pi^{2}} \bar{s}_{L}\sigma_{\mu\nu} F^{\mu\nu} b_{R}$,
$O_{g} = \frac{g_{s}}{16\pi^{2}} \bar{s}_{L} \sigma_{\mu\nu} G^{\mu\nu} b_{R}$,
and the current-current operator,  $O_{2} = \bar{s}_{L} \gamma_{\mu} q_{L}\bar{q}_{L}\gamma^{\mu}b_{L}$,
respectively.  $C^{R}_{7}$ and $C^{R}_{g}$
are the corresponding coefficients of the non-standard dipole operators involving 
right-handed light-quark fields, which appear in supersymmetric theories. 
We will use the numerical values of the
coefficients $a_{ij}$ as computed using the parton model 
in Ref.~\cite{Kagan:1998bh}:
$a_{27}\approx 0.0123$, $a_{g7}\approx -0.0952$ and $a_{2g}\approx 0.0010$.
In order to explore the implications of supersymmetric flavor models it is 
useful to express the effective coefficients in terms of the new physics contributions
to the Wilson coefficients at the scale $m_{W}$. To this end 
numerical expressions were given in Ref.~\cite{Kagan:1998bh}
including NLO renormalization effects from $m_{W}$ down to the $m_{b}$ mass scale,
\begin{eqnarray}
C_{7} &=& C_{7}^{0} +\eta_{77}~C_{7}(m_{W})+\eta_{7g} ~C_{g}(m_{W}), \\
C_{g} &=& C_{g}^{0} +\eta_{g}~C_{g}(m_{W}).
\label{Ceffec}
\end{eqnarray}
Here $\eta_{77}=0.67$, $\eta_{7g}=0.09$
and $\eta_{g}=0.70$. The supersymmetric contributions to 
$C^{L}_{g}$ were given in Eq.~\ref{CgBphiKs}.
$C_{7}^{L}(m_{\widetilde{q}})$ is given by, 
\begin{equation}
C^{L}_{7}(m_{\widetilde{d}}) = 
\frac{\alpha_{s} \pi}{ 2 m_{\widetilde{b}}} 
\left[
\frac{m_{b}}{3 m_{\widetilde{b}}}
M_{3}(x) (\delta_{\rm LL}^{d})_{23} +\frac{1}{10}
M_{1}(x) (\delta^{d}_{\rm LR})_{23}
\right],
\label{C7BphiKs}
\end{equation}
where $M_{1}(x)$ and $M_{3}(x)$ are dimensionless form factors
defined in Eqs.~\ref{M1x} and \ref{M3x} of the appendix.
$M_{1}(x)$ and $M_{3}(x)$ have been normalized to 1 when $x\rightarrow 1$.
We note that for simplicity we have defined $C_{7}$ as the whole coefficient
accompanying the operator $O_{7}$. Therefore in our notation
the SM contribution to the Wilson coefficient $C_{7}^{L}$
at $m_{W}$ is given by 
$ C^{\rm SM}_{7} (m_{W}) = 
- \sqrt{2} m_{b} G_{F} V_{ts}^{*} V_{tb} K \left( x_{t} \right)$,
where  $x_{t}= m_{t} / m_{W}$ and
$K(x)$ is a dimensionless form factor given in Eq.~\ref{Kx}
of the appendix.
The supersymmetric 
contributions to $C_{2}$ are neglibible. We will use the SM value, 
$C_{2}(m_{b}) \approx 1.11 \times G_{F} V_{ts}^{*} V_{tb} /\sqrt{2}$. 
It is straigthforward to obtain $C^{R}_{g}$ and $C_{7}^{R}$ 
by the exchange $L\leftrightarrow R$ in the expressions for 
$C_{7}^{L}$ and $C_{g}^{L}$. 
Barring cancellations between $\delta_{\rm LR}$ and $\delta_{\rm LL}$ 
terms we will obtain an approximate bound
from the LR contribution. We can see from Eq.~\ref{AXsgamma} that the total $\delta_{\rm LR}$ contribution
is proportional to a coupling of the form 
$((\delta_{\rm LR})^{*}_{23} +(\delta_{\rm LR})^{*}_{32})$. 
We obtain the following approximate expression for the 
asymmetry,
\begin{equation}
\left. A_{\rm CP}^{b\rightarrow s \gamma} \right|_{\rm LR+RL}
 \approx (-)
\frac{\alpha_{s}\pi}{2 m_{\widetilde{b}}}\frac{C_{2} {\rm Im} [ (\delta_{\rm LR})^{*}_{23} +(\delta_{\rm LR})^{*}_{32} ]
A(x)}{\left|C^{\rm SM}_{7}\right|^{2}}
\end{equation}
For the $\delta_{LL,RR}$ couplings we will obtain a similar
expression proportional to the coupling $((\delta_{\rm LL})^{*}_{23} +(\delta_{\rm RR})^{*}_{32})$.
Here $A(x)$ is a dimensionless form factor defined by, 
\begin{eqnarray}
A(x) &=&  (a_{27} \eta_{77}\frac{1}{3} M_{1}(x) + 
\nonumber
\\ 
&&a_{2g} (\eta_{7g}\frac{1}{3} M_{1}(x) + \eta_{g} M(x) ))
 \end{eqnarray}
For the model under consideration, 
\begin{equation}
(\delta^{d}_{\rm LR})^{*}_{23} + (\delta^{d}_{\rm LR})^{*}_{32} = 
 (-) 4 \lambda s_{\phi} \omega \frac{v A_{b}}{m^{2}_{\widetilde{b}}}c_{\beta}.
\label{dLRLR23}
\end{equation}
The SM contribution to the effective coefficient, $C_{7}^{0}$ is related
with the Wilson coefficient at the $m_{W}$ scale 
by a renormalization
factor, $C_{7}^{0} = \eta_{bW} C_{7}^{\rm SM}$, 
which can be extracted from Ref.~\cite{Kagan:1998bh}.
Assuming
that $x\approx 1$, {\it i.e.} $m_{\widetilde{g}} \approx  m_{\widetilde{b}}$,
and using the values of $\lambda$ and $\omega$ as determined
from quark masses and mixings we obtain the constraint, 
\begin{equation}
\left. A_{\rm CP}^{b\rightarrow s \gamma} \right|_{\rm LR+RL}
 \approx
s_{\phi} c_{\beta}  
\left(  \frac{40 \rm ~GeV}{m_{\widetilde{b}}} \right)^{2}
\left( \frac{A_{b}}{m_{\widetilde{d}}} \right) \lesssim 0.1
\end{equation}
In the worst case scenario, 
assuming that $s_{\phi} \approx 1$, $A_{b}\approx m_{\widetilde{d}}$,
$\tan\beta \approx 1$
the current experimental bound requires
$m_{\widetilde{b}} \gtrsim 230$~GeV.
On the other hand, for large $\tan\beta$
one would obtain a milder constraint. We would like to point out that
the phase $\phi$ is not constrained by
the CKM phase. If $s_{\phi} \ll 1$ 
the squark masses would not be constrainted by $A_{\rm CP}^{b\rightarrow s \gamma}$.
One would naively expect that 
the $\delta_{\rm LR}$ gives the
dominant contribution 
to $A_{\rm CP}^{b\rightarrow s \gamma}$ because of the $m_{b}/v$ suppression 
factor of the $\delta_{\rm LL}$ contributions to $C_{7}$ and $C_{g}$. 
Nevertheless for the model under consideration the contributions coming
from $\delta_{\rm LL,RR}$ couplings are of the same order of magnitude.
For the model under consideration, 
\begin{equation}
(\delta^{d}_{\rm LL})^{*}_{23} + (\delta^{d}_{\rm RR})^{*}_{32} = 
- 4 \lambda \rho^{\prime} e^{-i\phi}.
\end{equation}
We obtain a similar expression, 
\begin{equation}
\left. A_{\rm CP}^{b\rightarrow s \gamma} \right|_{\rm LL+RR}
 \approx
s_{\phi}  \rho^{\prime}
\left(  \frac{30 \rm ~GeV}{m_{\widetilde{b}}} \right)^{2}
 \lesssim 0.1.
\end{equation}
The constraint on the squark spectra depends on the value of $\rho^{\prime}$.
For the large $\tan\beta$ case we noted in Sec.~\ref{Radmasses} that
$\rho=\widetilde{m}^{2}/m_{\widetilde{b}}^{2}\approx 9$. 
If $\rho^{\prime} =\rho$ we would obtain the constraint 
$m_{\widetilde{b}} \gtrsim 300$~GeV. This constraint could be avoided if
$\phi \approx 0$ or $\eta^{\prime}\ll 1$.
\subsection{$\Gamma(b\rightarrow s \gamma)$ }
The supersymmetric contributions to the $b \rightarrow s \gamma$ decay 
are indirectly correlated with the CP asymmetries in 
$B\rightarrow \phi K_{s}$ and $b\rightarrow s\gamma$ decays
since the same flavor mixing couplings contribute to the relevant Wilson coefficients.
The $b\rightarrow s \gamma$ decay rate is also 
proportional to the $C_{7}$ Wilson coefficients, 
\begin{equation}
\Gamma (b\rightarrow s \gamma) \propto 
(\left| C^{L}_{7} \right|^{2}
+ \left| C^{R}_{7} \right|^{2}).
\label{bsgdecay}
\end{equation}
The current world average
of the CLEO \cite{Chen:2001fj} and BELLE \cite{Abe:2001hk} results
is given by $B(b\rightarrow s \gamma)_{\rm exp} = (3.3\pm 0.4) \times 10^{-4}$,
which can perfectly be accounted for the SM theoretical prediction,
$B(b\rightarrow s \gamma)_{\rm theo} = (3.29\pm 0.33) \times 10^{-4}$ \cite{SMbsgam},
which leaves a small window open for new physics.
There is no SM contribution to $C_{7}^{R}$.
A full expression for the main supersymmetric contributions,
i.e. gluino exchange, to this branching ratio were first
given in Ref.~\cite{Bertolini:1986tg}.
Further improvements in the calculation as
chargino diagrams and QCD corrections
were subsequently included \cite{morebsgam}.
Therefore if the
supersymmetric contribution is just a correction to the SM one 
we can expand in powers of $R_{s\gamma} = 
\left|C^{L~ \rm SUSY}_{7} / C^{\rm SM}_{7}\right|$
and obtain to leading order,
\begin{equation}
\frac{ \Gamma (b\rightarrow s \gamma)}{ \Gamma (b\rightarrow s \gamma)^{\rm SM}} -1
\approx  2 R_{s\gamma}.
\label{bsgara}
\end{equation}
Allowing the supersymmetric contribution to saturate 
the $2\sigma$ experimental uncertainity 
we obtain to leading order in $R_{s\gamma} $ the constraint,
\begin{equation}
  R_{s\gamma} \lesssim \frac{  \Delta ( \left. {\rm Br} (b\rightarrow s \gamma)  \right|_{\rm exp})}{ 
\left.  {\rm Br} (b\rightarrow s \gamma)  \right|_{\rm exp}} \lesssim 0.12,  
\label{bsgcons}
\end{equation}
Using the expression for the supersymmetric contribution to $C_{7}^{L}$ 
from Eq.~\ref{C7BphiKs} we obtain for the $\delta_{\rm LR}$ 
contribution to $R_{s\gamma}$ the expression,
\begin{equation}
\left. R_{s\gamma} \right|_{\rm LR}  =
\frac{\eta_{s\gamma}}{m_{\tilde{q}}} \left| (\delta^{d}_{LR})_{23}\right|  M_{3}(x) .
\label{RbsgLR}
\end{equation}
Here $\eta_{s\gamma}$ is a coefficient independent of the supersymmetric parameter space
with mass units. Using the SM expression for 
$C_{7}^{L}$ and 
the expression for $(\delta^{d}_{\rm LR})_{23}$
that our model predicts for $x\approx 1$ , 
$$
\left|(\delta^{d}_{\rm LR})_{23}\right| \approx  2 \lambda c_{\beta}
 \frac{v}{m_{\widetilde{b}}}
  \frac{\widetilde{m}}{m_{\widetilde{b}}},
$$
we obtain the constraint, 
\begin{equation}
c_{\beta} \left( \frac{130~{\rm GeV}}{m_{\widetilde{b}}} \right)^{2}
\left( \frac{\widetilde{m}}{m_{\widetilde{b}}} \right) \lesssim 0.12.
\label{Rbsgcons}
\end{equation}
For the large $\tan\beta$ case with $m_{\widetilde{g}} \approx m_{\widetilde{b}}$
examined in Sec.~\ref{Radmasses}
, $\tan\beta=50$, $\widetilde{m}$ was 
required to be $\widetilde{m}\approx 3 m_{\widetilde{b}}$. In this case
we obtain the following constraint on the squark mass scale,
$m_{\widetilde{b}}\gtrsim 370$~GeV.
An analysis with similar results 
can be implemented for the $\delta_{\rm (LL+RR)}$ contribution to $\Gamma(b\rightarrow s \gamma)$.
In this case,
$$
\left|(\delta^{d}_{\rm LL})_{23}\right| \approx  2 \lambda \rho^{\prime}.
$$
We obtain the constraint, 
\begin{equation}
\rho^{\prime} \left( \frac{16~{\rm GeV}}{m_{\widetilde{b}}} \right)^{2}
\lesssim 0.12.
\label{Rbsgcons}
\end{equation}
For the large $\tan\beta$ case with $m_{\widetilde{g}} \approx m_{\widetilde{b}}$
examined in Sec.~\ref{Radmasses}
, $\tan\beta=50$, $\rho$ was 
required to be $\rho \approx 3$. If $\rho^{\prime}=\rho$
we would obtain the following lower bound on the squark mass scale,
$m_{\widetilde{b}}\gtrsim 140$~GeV.
\section{Conclusions}
We have shown that generic supersymmetric flavor models exist for the
radiative generation of fermion masses, mixings
and CP phases. 
We have studied in detail the phenomenological
implications of a particular supersymmetric flavor model 
for the radiative generation of first and second generation quark masses, 
focusing our attention especially in the CP violating phenomenology.
The basic idea underlying this kind of flavor models
is that the flavor breaking fields are also supersymmetry breaking fields.

We have shown that these models generically solve the SUSY CP problem 
in a very simple fashion.
The one-loop contributions to EDMs are exactly zero due to the real
character of the relevant parameters while the two-loop contributions
are suppressed. 

Our main goal was to present a flavor model as predictive as possible. 
To this end we have proposed a particular implementation of this scenario 
using a U(2) flavor symmetry where the required hierarchy of
flavor breaking vevs is expressed in powers of a single parameter,
$\lambda$. As a consequence the model generates quark Yukawa matrices that 
contain only three parameters, $\lambda,\theta,\gamma$ and can fit the data with
precision. Therefore the quark masses and mixings determine the amount
of flavor violation in the soft sector requiring a very heavy susy espectra especially
to overcome the constraints on $\Delta m_{K}$, $\epsilon$ and $\Delta m_{d}$. 

We would like to emphasize that this case study can be considered the 
worst case scenario from the point of view of FCNC constraints. 
Between the extreme case study here considered and the usual 
models with scalar flavor breaking vevs there is a continua of possibilities
which would ameliorate the FCNC constraints. For instance, we could increase
the number of parameters in the flavor breaking sector, use a different
flavor symmetry or generate radiatively only the first generation of fermion masses.
If that was the case one could lower considerably the constraints 
on the sfermion spectra, probably at the price of decreasing 
the predictivity of the flavor model.

We believe these models 
are an scenario worth of a more dedicated exploration.
They generically allow us to simplify the ``flavor vacuum'', or in other words 
the hierarchies of the flavor breaking vevs, through the introduction of a natural
hierarchy, the loop factor, and they offer a new insight in the SUSY CP and flavor problems.
\section{Appendix}
For completeness we include expressions for 
the dimensionless form factors that were used in the main text.
The form factor $F(x,y,z)$ is defined as,
\begin{equation}
F(x,y,z) = \frac{\left[ (x^{2}y^{2} \ln \frac{y^{2}}{x^{2}} + y^{2}z^{2} \ln \frac{z^{2}}{y^{2}}
+ z^{2}x^{2} \ln \frac{x^{2}}{z^{2}} \right]}{(x^{2}-y^{2})(y^{2}-z^{2})(z^{2}-x^{2})}.
\label{Ffun}
\end{equation}
$f(x)$, $g(x)$ and $h(x)$ are given by,
\begin{eqnarray}
f(x) &=& \frac{10x }{3}\left[\frac{(x^{2} -8x -17)}{(x-1)^{4}} + 
 \frac{6(1+3x) \ln(x))}{(x-1)^{5}}\right], ~~~
\label{fx} \\
g(x) &=& 10 \left[ \frac{(x^{2} + 10x +1)}{(x-1)^{4}} - \frac{ 6 x (1+x) \ln(x))}{(x-1)^{5}}\right], 
\label{gx} \\
h(x) &=&  \frac{11 g(x)- 6f(x)}{5}. 
\label{hx} 
\end{eqnarray}
These functions appear in the calculation of the supersymmetric
contributions to the Wilson coefficients 
The functions $f(x)$, $g(x)$ and $h(x)$ have been conveniently normalized so that 
in the limit $x\rightarrow 1$ they tend to 1. Approximate expressions
in the limits $x\rightarrow 0,1$ are given by,
\begin{equation}
f(x) = \left\{
\begin{array}{c}
1 - \frac{1}{3} \zeta + {\cal O}(\zeta^{2}) , \quad x \rightarrow 1 , (\zeta=x-1) \\
- \frac{10 x}{3} (17 + 6 \ln (x) )  + {\cal O}(x^{2}) , \quad x \rightarrow 0,
\end{array}
\right.
\end{equation}
\begin{equation}
g(x) = \left\{
\begin{array}{c}
 1 - \zeta + {\cal O}(\zeta^{2}) , \quad x \rightarrow 1 , (\zeta=x-1) \\
10 ( 1 + 2 x (3 \ln (x) + 7 )  + {\cal O}(x^{2}) , \quad x \rightarrow 0,
\end{array}
\right.
\end{equation}
$N(x)$ and $M(x)$ 
are dimensionless form factors given by,
\begin{eqnarray}
N(x) &=& \left[ \frac{(x^{2} +172 x +19)}{36 (x-1)^{4}} \right.
\nonumber
\\
&& \left. - x\ln (x) \frac{( x^{2} -15 x -18)}{6 (x-1)^{5}} \right] 
\label{Nxform} 
\\
M(x) &=& \sqrt{x} \left[ \frac{(54 x^{4} +216 x^{3} -287 x^{2} -8 x +1)}{9 (x-1)^{4}}
\right.
\nonumber
 \\
&& \left. -  2 x^{2} \ln (x) \frac{(36 x^{2} -19 x -21)}{3 (x-1)^{5}} \right]
\label{Mxform}
 \end{eqnarray}
In the limit $x\simeq 1$ $M(x)$ is given by,
\begin{equation}
M(x) = a_{0} + a_{1} (x-1) + {\cal O}((x-1)^{2}),\quad 
a_{0} = \frac{31}{30}, ~a_{1} = \frac{233}{180} .
\label{Mx1}
\end{equation}
If one is interested in the limit 
$x\simeq 0$, {\it i.e.} $m_{\tilde{g}}\ll m_{\tilde{q}}$,
it is also possible to obtain an approximate expression for $M(x)$,
\begin{equation}
M(x) = \frac{1}{9} \sqrt{x} \left( b_{0} + b_{1} x + {\cal O}(x^{2}),\right) \quad 
b_{0} = 1, ~ b_{1} = -4 .
\label{Mx0}
\end{equation}
The functions $M_{1}(x)$ and $M_{3}(x)$ are defined by, 
\begin{eqnarray}
M_{1}(x) &=&  \frac{12 x^{2} \ln(x)}{(1-x)^{4}} + \frac{6(1+5x)}{(1-x)^{3}}
\label{M1x}
\\
M_{3}(x) &=& 
\frac{10 \sqrt{x}}{3}\left[  \frac{(1 - 8x   -17x^{2} )}{(x-1)^{4}} + 
 \ln(x) \frac{(18 x^{2}  + 6 x^{3})}{(x-1)^{5}} \right].~~~~
\label{M3x}
 \end{eqnarray}
Here $M_{1}(x)$ and $M_{3}(x)$ have also been ``normalized'' so 
that in the limit $x\rightarrow 1$ they tend to 1. Finally
the dimensionless form factor K(x), which appears in the 
SM contribution to the Wilson coefficient $C_{7}$, is given by,
\begin{equation}
K(x) =  \frac{x}{2(x-1)} \left( \frac{(8 x^{2} + 5 x - 7)}{12} - 
\left( \frac{2x^{2}}{3} -x \right) \frac{\ln (x) }{(x-1)} \right) 
\label{Kx}
\end{equation}
\acknowledgements
One of us (J.F.) is grateful to Michelle Papucci and Gilad Perez for suggestions
regarding the calculation of CP asymmetries in B decays. 
This work is supported by
the Ministry of Science of Spain under grant EX2004-0238,
the US Department of Energy under Contracts DE-AC03-76SF00098 and 
DE-FG03-91ER-40676, and
by the National Science Foundation under grant PHY-0098840.
The work of J.L.D.-C. is supported by CONACYT-SNi (Mexico).



\begin{references}
\bibitem{EDMscomment}
At present the neutron EDM, $d_{n}$, 
is constrained to be $d_{n}<6.3\times 10^{-26}$~e~cm.
and the mercury atom EDM, $d_{H_{g}}$, 
is constrained to be $d_{H_{g}}<2.1\times 10^{-28}$~e~cm.
It is expected that current limits will be improved by one to six orders
of magnitude.
\end{references}
\end{document}